\documentclass[aps,prl,twocolumn,superscriptaddress]{revtex4-2}

\usepackage{amsmath,amssymb,braket}
\usepackage{amsfonts}
\usepackage{graphicx}
\usepackage{color}
\usepackage{bm}
\usepackage[dvipsnames]{xcolor}
\usepackage{graphicx}
\usepackage{hyperref} 
\usepackage{cleveref} 
\newcommand{\sectionprl}[1]{{\par\it #1.---}}

\usepackage[compat=1.1.0]{tikz-feynman}
\usetikzlibrary{decorations.pathmorphing,calc}

\tikzset{
  wavy-big/.style={
    line width=1pt,
    line cap=round,
    decorate,
    decoration={snake, amplitude=2.2pt, segment length=7pt}
  }
}

\newcommand{\VCsetup}{%
  \useasboundingbox (-2.2,-3.6) rectangle (8.6,3.6); 
  \begin{feynman}
    \vertex (Lext) at (-2,0);
    \vertex (L)    at (0,0);
    \vertex (T)    at (3,  1.732);
    \vertex (B)    at (3, -1.732);

    \vertex (Rt)   at (6,  3.464);
    \vertex (R)    at (6,  0);
    \vertex (Rb)   at (6, -3.464);

    \vertex (RextT) at (8.2,  4.755);
    \vertex (RextB) at (8.2, -4.755);

    \diagram*{ (Lext) -- [wavy-big] (L) };
    \draw (L) -- (T);
    \draw (L) -- (B);

    \draw (RextT) -- (Rt);
    \draw (RextB) -- (Rb);
}

\newcommand{\VCA}{%
\begin{tikzpicture}
  \VCsetup
    \draw[wavy-big] (Rt) -- (T);  \draw (Rt) -- (R);
    \draw[wavy-big] (Rb) -- (B);  \draw (Rb) -- (R);
  \end{feynman}

  \node[circle, fill, inner sep=2pt] at (L)  {};
  \node[circle, fill, inner sep=2pt] at (Rt) {};
  \node[circle, fill, inner sep=2pt] at (Rb) {};

  \node[left, yshift=10pt, font=\Huge]  at ($(Lext)!0.5!(L)$) {$\alpha$};
  \node[right, yshift=-10pt, font=\Huge]  at ($(RextT)!0.5!(Rt)$) {$\beta$};
  \node[right,yshift=10pt, font=\Huge]  at ($(RextB)!0.5!(Rb)$) {$\gamma$};

  \node[above, yshift=5pt, font=\Huge]  at ($(L)!0.5!(T)$) {$\mu$};
  \node[right, font=\Huge]  at (R)              {$\nu$};
  \node[below,yshift=-5pt, font=\Huge]  at ($(L)!0.5!(B)$)  {$\rho$};
\end{tikzpicture}%
}

\newcommand{\VCB}{%
\begin{tikzpicture}
  \VCsetup
    \draw[wavy-big] (Rt) -- (T);  \draw (Rt) -- (R);          
    \draw          (Rb) -- (B);  \draw[wavy-big] (Rb) -- (R); 
  \end{feynman}

  \node[circle, fill, inner sep=2pt] at (L)  {};
  \node[circle, fill, inner sep=2pt] at (Rt) {};
  \node[circle, fill, inner sep=2pt] at (Rb) {};

  \node[left, yshift=10pt, font=\Huge]  at ($(Lext)!0.5!(L)$) {$\alpha$};
  \node[right, yshift=-10pt, font=\Huge]  at ($(RextT)!0.5!(Rt)$) {$\beta$};
  \node[right,yshift=10pt, font=\Huge]  at ($(RextB)!0.5!(Rb)$) {$\gamma$};

  \node[above,yshift=5pt, font=\Huge]  at ($(L)!0.5!(T)$) {$\mu$};
  \node[right, font=\Huge]  at (R)              {$\nu$};
  \node[below,yshift=-5pt, font=\Huge]  at ($(L)!0.5!(B)$)  {$\rho$};
\end{tikzpicture}%
}

\newcommand{\VCC}{%
\begin{tikzpicture}
  \VCsetup
    \draw         (Rt) -- (T);   \draw[wavy-big] (Rt) -- (R); 
    \draw[wavy-big] (Rb) -- (B); \draw (Rb) -- (R);           
  \end{feynman}

  \node[circle, fill, inner sep=2pt] at (L)  {};
  \node[circle, fill, inner sep=2pt] at (Rt) {};
  \node[circle, fill, inner sep=2pt] at (Rb) {};

  \node[left, yshift=10pt, font=\Huge]  at ($(Lext)!0.5!(L)$) {$\alpha$};
  \node[right, yshift=-10pt, font=\Huge]  at ($(RextT)!0.5!(Rt)$) {$\beta$};
  \node[right,yshift=10pt, font=\Huge]  at ($(RextB)!0.5!(Rb)$) {$\gamma$};

  \node[above, yshift=5pt,font=\Huge]  at ($(L)!0.5!(T)$) {$\mu$};
  \node[right, font=\Huge]  at (R)              {$\nu$};
  \node[below, yshift=-5pt,font=\Huge]  at ($(L)!0.5!(B)$)  {$\rho$};
\end{tikzpicture}%
}

\usepackage{braket}
\usepackage{comment}
\usepackage{color}
\usepackage{tabularx}
\newcolumntype{C}{>{\centering\arraybackslash}X}
\newcolumntype{L}{>{\raggedright\arraybackslash}X}
\newcolumntype{R}{>{\raggedleft\arraybackslash}X}

\newcommand{\cs}{c_s}
\newcommand{\tcs}{\tilde{c}_s}
\newcommand{\tD}{\tilde{D}}
\newcommand{\tl}{\tilde{\lambda}}
\newcommand{\Gf}{g}
\newcommand{\Co}{C}
\newcommand{\Cn}{\tilde{C}}
\newcommand{\Sf}{\mathbb{S}}
\newcommand{\bSf}{s}
\newcommand{\Pfp}{\mathbb{P}}
\newcommand{\iCH}{\hat{H}}
\newcommand{\Hgll}{H^2_{11}}
\newcommand{\Hggg}{H^2_{22}}
\newcommand{\const}{\mathcal{C}}
\newcommand{\anoexp}{\eta}

\usepackage[normalem]{ulem}
\usepackage{xcolor}

\begin{document}

\title{ 
Symmetry-based nonlinear fluctuating hydrodynamics in one dimension
}

\author{Yuki Minami}
\affiliation{Faculty of Engineering, Gifu University, Yanagido, Gifu 501-1193, Japan}
\email{minami.yuki.u5@f.gifu-u.ac.jp}
\author{Hiroyoshi Nakano}
\affiliation{Institute for Solid State Physics, University of Tokyo, 5-1-5, Kashiwanoha, Kashiwa 277-8581, Japan}
\email{nakano.hiroyoshi.7n@issp.u-tokyo.ac.jp}
\author{Keiji Saito}
\affiliation{Department of Physics, Kyoto University, Kyoto 606-8502, Japan}
\email{saitoh@scphys.kyoto-u.ac.jp}

\date{\today}

\begin{abstract}
We present a symmetry-based formulation of nonlinear fluctuating hydrodynamics (NFH) for one-dimensional many-particle systems with generic homogeneous nearest-neighbor interactions. We derive the hydrodynamic equations solely from symmetry and conservation principles, ensuring full consistency with  thermalization. Using the dynamic renormalization group, we 
identify a KPZ-type fixed point, characterized by the dynamical exponent $z=3/2$ for both the sound and heat modes.
Extensive numerical simulations of the derived NFH equations confirm this exponent and further reveal that both modes are  close to the universal KPZ scaling function—the Pr\"{a}hofer–Spohn function.
These findings establish a unified, symmetry-based framework for understanding universal transport and fluctuation phenomena in one-dimensional nonequilibrium systems, independent of microscopic details.
\end{abstract}

\maketitle

\sectionprl{Introduction}
Recently, hydrodynamic theory has emerged as a unifying framework for understanding a broad range of transport phenomena in solids, encompassing both electronic and thermal conduction~\cite{bandurin2016negative,gooth2018thermal,moll2016evidence,krishna2017superballistic,lee2015hydrodynamic,sulpizio2019visualizing}. 
Direct evidence of viscous electronic fluids has been observed~\cite{bandurin2016negative,gooth2018thermal}, and Poiseuille-like electron flow has been reported in both two- and three-dimensional systems~\cite{moll2016evidence,krishna2017superballistic,lee2015hydrodynamic,sulpizio2019visualizing}. 
On the theoretical side, substantial progress has been made in one-dimensional (1D) systems, where the hydrodynamics of integrable models has been systematically derived~\cite{doyon,rcdd,gge}, followed by experimental corroboration~\cite{ghdexp}. 
For nonintegrable 1D many-particle Hamiltonians, the framework of nonlinear fluctuating hydrodynamics (NFH) has been proposed~\cite{beijeren2012,spohn2014nonlinear,spohn2016fluctuating}, attracting considerable attention~\cite{das2014,kulkarni15,saito2021microscopic}.

The NFH theory was originally developed as a semi-phenomenological approach for many-particle systems with generic homogeneous nearest-neighbor interactions, based on the assumptions of local thermal equilibrium and perturbative calculations~\cite{spohn2014nonlinear}; its microscopic foundations have since been explored~\cite{saito2021microscopic}. Analogous to  conventional hydrodynamics, the NFH equations are formulated in terms of three fundamental conserved quantities: the local stretch, momentum, and energy, where the stretch corresponds to the conservation of total length. These conserved fields decompose into three physically distinct collective modes—two counterpropagating sound modes and a nonpropagating heat mode.
Mode-coupling analysis has predicted that the sound modes belong to the Kardar–Parisi–Zhang (KPZ) universality class, whereas the heat mode exhibits L\'{e}vy-type diffusion~\cite{spohn2014nonlinear,spohn2016fluctuating}.
Note that the NFH equations themselves have not yet been numerically studied.
This pioneering framework represents a major milestone in elucidating hydrodynamic behavior in interacting isolated systems and has stimulated extensive subsequent research \cite{das2014,kulkarni15,de2023nonlinear,saito2021microscopic,hiura2023microscopic,popkov2015fibonacci,spohn2015nonlinear,das2020nonlinear,lepri2020too,gopalakrishnan2024non,nishikawa2025energy}. However, the original hydrodynamic equations are known to exhibit certain theoretical inconsistencies.  In particular, the nonlinear coupling coefficients must take unphysical values when the symmetry associated with thermal equilibrium is imposed (see Sec. IX of the Supplemental Material (SM)~\cite{SM1}). 

To deepen our understanding of the hydrodynamic picture, it is therefore both natural and desirable to seek an alternative formulation.
We further note that, whereas the fluctuating Navier–Stokes equations have long been subjected to dynamic renormalization-group (RG) analysis since the seminal work of Forster, Nelson, and Stephen~\cite{fns,smith1998renormalization}, 
no comparable RG analysis has yet been developed for NFH. 
Likewise, while direct numerical simulations of the fluctuating Navier–Stokes equations are now common in studies of turbulence~\cite{Bell2022-jo, Ishan2025-fw, Gosteva2025-cq} and long-range correlations~\cite{Donev2014-jy, Srivastava2023-nx},
they have remained largely unexplored in the context of NFH. Against this background, we introduce a symmetry-consistent revision of the NFH equations and perform both RG and numerical analyses.

In physics, numerous successful predictions have been made by considering symmetry alone to determine how macroscopic properties are constrained. Hydrodynamics can be understood as a gradient expansion for currents of conserved quantities \cite{landau1987fluid,polchinski1992effective, kaplan2005five, Liu:2018Uz}.
Following this viewpoint, we formulate the most general hydrodynamic theory by imposing only symmetry, conservation laws, and equilibrium conditions within the gradient-expansion framework, without relying on microscopic Hamiltonian details.
We then construct symmetry-based NFH equations up to the minimal orders set by scaling arguments [Eqs.~(\ref{EOM original main}) with (\ref{A matrix})–(\ref{H3 matrix main}), and Eqs.~(\ref{EOM phi main}) with (\ref{cs, D_0, D_s})–(\ref{parametrized Gmatrix})].
We perform a dynamic RG analysis of these equations  and identify a KPZ-type fixed point, characterized by a dynamical exponent $z=3/2$ for both the sound and heat modes.
The exponent of the heat mode differs from earlier theoretical expectations~\cite{spohn2014nonlinear}. These exponents should be directly related to the anomalous heat transport observed in one-dimensional systems~\cite{lepri2016book,benenti2023non,Chang2008-zw,Xu2014-oo,Lee2017-gq}. Furthermore, our extensive numerical simulations not only confirm the dynamical exponent $z=3/2$, but also show that both the heat and sound modes exhibit space-time correlations that are close to the universal KPZ scaling function, i.e., the Pr\"{a}hofer–Spohn function~\cite{prahofer2004exact}.
These findings establish a symmetry-based hydrodynamic framework that accounts for universal scaling behaviors and fluctuation dynamics, independent of microscopic details.

\sectionprl{Symmetry-based NFH equations for the general systems}
We consider an $N$-particle system in one dimension, described by the Hamiltonian
\begin{align}
H &= \sum_{i} \frac{1}{2} p_i^2 +  \sum_{i } V(q_{i+1} - q_i),
\end{align} 
where $p_i$ and $q_i$ are the momentum and position of the $i$-th particle, respectively. Here, $i \in {\mathbb Z}/N {\mathbb Z}$, and we consider infinite line picture by imposing the condition $q_{i+N}= q_{i} + L$ with the total length $L$ and $p_{i+N}=p_i$. The potential $V$ is a general nonlinear interaction between nearest-neighbor sites. Examples include the Fermi-Pasta-Ulam-Tsingou potential \cite{mendl2013dynamic} and a hard-shoulder potential \cite{mendl2014equilibrium}. In the main part of this paper, we consider a general system without inversion symmetry, i.e., $V(-x) \neq V(x)$. We discuss the inversion-symmetric case at the end of this paper.

In this setup, the conserved quantities are the system size, total momentum, and total energy. Through a relevant coarse-graining procedure in space, one can introduce local conserved densities that depend on the continuous variables $x$ and $t$. Let $l(x,t)$, $g(x,t)$, and $e(x,t)$ be the stretch, momentum-density, and energy-density fields, which are local fields corresponding to the system size, total momentum, and total energy, respectively.
 We consider their deviations from the equilibrium values, which we collectively denote by $\bm{u} = (\delta l, \delta g, \delta e)$.
 The local conserved densities are connected to corresponding local currents via the continuity equation. Through the projection of the full dynamics onto that of the conserved quantities, the fluctuating hydrodynamics can be formally derived \cite{zwanzig,saito2021microscopic,zubarevmorozof}. 
In general, the currents are functionals of $(u_a, \partial_x)$~\cite{polchinski1992effective, kaplan2005five, Liu:2018Uz}, and hence one can formally perform a gradient expansion, yielding
\begin{align}
\begin{split}
&\partial_t u_a + \partial_x \!\bigl( {\cal J}_{a} + {\cal J}^{\mathrm{ran}}_a \bigr) = 0, \\
&{\cal J}_{a} = \sum_b A_{ab} u_b - D_{ab}\, \partial_x u_b + \sum_{b,c} H^a_{bc}\, u_b u_c,\\
&\langle {\cal J}^{\mathrm{ran}}_a(x,t){\cal J}^{\mathrm{ran}}_b(x',t') \rangle=2B_{ab}\delta(t-t')\delta(x-x'),
\label{EOM original main}
\end{split}
\end{align}
where $A_{ab}$, $D_{ab}$, and $H^a_{bc}$ are formal expansion coefficients. 
The variables ${\cal J}^{\mathrm{ran}}_a$ are noise terms arising from the projection of the total phase-space dynamics onto the conserved quantities, and $B_{ab}$ assumed to be connected to the matrix $D$ through the fluctuation–dissipation relation~\cite{landau1987fluid,kth}.
In Eq.~(\ref{EOM original main}), the expansion is systematically truncated at the relevant order set by scaling arguments (see Sec.~IV of the SM~\cite{SM1}). 

The formal expansion coefficients are determined solely from symmetry and conservation-law considerations. From the local conservation of mass~\cite{saito2021microscopic}, one obtains $\partial_t u_1 - \partial_x u_2 = 0$, which leads to the conditions $H^a_{bc} = D_{ab} = 0$ for $a = 1$. We next consider the time-reversal symmetry and a general potential lacking inversion symmetry, $V(-x) \neq V(x)$.
In addition, we require thermal equilibrium as the steady state, which implies that the steady state distribution of $u$ obeys the Gaussian distribution at the equilibrium $P_{\mathrm{eq}}[{u}] =Z^{-1}\exp\bigl(-\frac{1}{2} \sum_{a,b} \int dx  u_a(x)  [\Co^{-1}]_{ab}  u_b(x)\bigr)$, where  $\Co$ is the covariance matrix and the indices $a$ and $b$ are taken to run over $1,2,3$. By applying these physical requirements, the coefficients are determined explicitly as Eqs.~(\ref{A matrix})-(\ref{H3 matrix main}) in Appendix A of the End Matter. See also Sec.~I-II of the SM for the detailed derivation \cite{SM1}. 

Eq.~(\ref{EOM original main}) is effectively converted into three modes in the long-wavelength regime \cite{spohn2014nonlinear, spohn2016fluctuating}. The new fields $\phi$ are defined as $\phi^{\alpha} = \sum_b R_{\alpha,b} u_b$ through the matrix $R$ diagonalizing  $A$ into $R A R^{-1}={\rm Diag}(c_s, 0 , -c_s)$. The values $\pm c_s$ represent the right (left) sound velocity. Accordingly, $\phi^{\alpha} (\alpha=\pm)$ are the right (left) sound modes, and $\phi^{0}$ is the heat mode. 
In this representation, the NFH equations are written in the form
\begin{align}
\begin{split}
( \partial_t + \alpha c_s\partial_x - D_{\alpha} &\partial_x^2  ) \phi^{\alpha}\! +\sum_{\beta, \gamma} G^{\alpha}_{\beta \gamma} \partial_x (\phi^{\beta} \phi^{\gamma}) = \zeta^{\alpha} \, , \label{EOM phi main} \\
\langle \zeta^{\alpha} (x,t) \zeta^{\beta} (x',t') \rangle &= -2D_{\alpha} \delta_{\alpha\beta} \partial_x^2 \delta (t-t') \delta (x-x') \, ,
\end{split}
\end{align}
where $\alpha=+,0,- $, and $\langle ... \rangle$ is the noise average. The constraints on the original coefficients in Eq.~(\ref{EOM original main})  translate into the following constraints on the parameters:
\begin{align}
\begin{split}
D_+&=D_-:=D_s \, , \\
G^\alpha_{\beta \gamma}&=G^\alpha_{\gamma \beta}, \quad
G^\alpha_{\beta \gamma}=-G^{-\alpha}_{-\beta -\gamma}, \quad 
G^\alpha_{\beta \gamma}
=G^\beta_{\gamma \alpha} \, . 
\end{split}
\end{align}
Here, $D_s$ and $D_0$ are the diffusion constants of the sound and heat modes, respectively. In what follows, we parameterize the independent nonzero elements of $G^\alpha_{\beta\gamma}$ as $\lambda_1=G^+_{00}$, $\lambda_2=G^+_{+0}$, $\lambda_3=G^-_{--}$, and $\lambda_4=G^+_{--}$. The explicit expressions of these coefficients are given in Eqs.~(\ref{cs, D_0, D_s})-(\ref{parametrized Gmatrix}) in Appendix A of the End Matter. 
The combination of Eq.~(\ref{EOM phi main}) and Eqs.~(\ref{cs, D_0, D_s})–(\ref{parametrized Gmatrix}) constitutes the basic form of the symmetry-based NFH equations, for which we perform the renormalization-group analysis and extensive numerical computations.
We note that the present derivation of hydrodynamics fully follows the standard logic of effective field theory, 
in which all symmetry-allowed relevant terms are systematically enumerated, independent of microscopic details. 
This approach has yielded many universal predictions across diverse areas of physics~\cite{polchinski1992effective,kaplan2005five,Liu:2018Uz}. 
The formulation of symmetry-based NFH constitutes the first main result of this paper.

\sectionprl{RG analysis}
To investigate the universal scaling behavior, we now perform a renormalization group (RG) analysis of the symmetry-based NFH. For the RG calculation, we employ the Martin–Siggia–Rose–Janssen–de~Dominicis (MSRJD) path-integral representation~\cite{martin1973statistical, janssen1976lagrangean, de1976technics, de1978dynamics}, where the dynamics (\ref{EOM phi main}) on the infinite line is written in the Fourier mode representation:
\begin{align}
    Z=& \int \prod_{\alpha=0,\pm}\mathcal{D} \phi^\alpha\mathcal{D}\pi^\alpha  
e^{-I[\phi,\pi]},\\
    I=&\sum_\alpha\int_k \frac{1}{2}
    \begin{pmatrix}
        \phi^\alpha_{-k} &\pi^\alpha_{-k}
    \end{pmatrix}
    \begin{pmatrix}
        0 & 1/\Gf_{-k}
        ^{\alpha} \\
 1/\Gf_k^{\alpha}& 2D_\alpha k^2
    \end{pmatrix}
    \begin{pmatrix}
        \phi^\alpha_{k} \\
        \pi^\alpha_{k}
    \end{pmatrix} \nonumber \\
&+\sum_{\alpha,\beta,\gamma}\int_{k_1}\int_{k_2}k_1 G^\alpha_{\beta\gamma}
 \pi^\alpha_{-k_1} \phi^{\beta}_{k_+}\phi^{\gamma}_{k_-} \, ,  \label{eq:Iint} \\
 \Gf_k^{\alpha}=&1/(\omega -\alpha\cs k+iD_\alpha k^2) \, . 
\end{align}
Here, the subscript $k$ in any function $f_k$ denotes that the function depends on $(k,\omega)$ where $k$ and $\omega$ denoting the wavenumber and frequency, respectively (hence, $\phi^\alpha_k=\phi^\alpha(k,\omega)$ and $\Gf_k^{\alpha}=\Gf^{\alpha}(k,\omega )$). The variable $\pi_k^\alpha(:=\pi^\alpha(k,\omega))$ is the auxiliary field. We have introduced the notation $\int_k:=\int^\Lambda_{-\Lambda}(dk/2\pi)\int^\infty_{-\infty}(d\omega/2\pi)$, 
$k_\pm:=k_1/2\pm k_2$, and $\omega_\pm:=\omega_1/2\pm \omega_2$, with $\Lambda$ being the wavenumber cutoff.

We perform the one-loop calculations to obtain the following RG equations (see the diagrams in Appendix B of the End Matter as well as the details in Sec.~V of the SM~\cite{SM1})
\begin{align}
            -\Lambda\frac{d D_s}{d \Lambda} =&D_s \biggl[\frac{\tl_1^2}{\tD}+\frac{4 \tl_2^2(\tD+1) }{\tcs^2+(\tD+1)^2}+\tl_3^2+\frac{\tl_4^2(\tcs^2+3)}{\tcs^2+1}\biggr], \label{RG eq Ds}\\
    -\Lambda\frac{d D_0}{d \Lambda} =&D_0 \biggl[\frac{8 (\tD+1) \tl_1^2}{\tcs^2+(\tD+1)^2}+2\tl_2^2\biggr],\\
              -\Lambda\frac{d \cs}{d \Lambda} =&
           -\Lambda\frac{d \lambda_i}{d \Lambda} =0 \qquad (i=1,2,3, 4),\label{RG eq c lambda}
\end{align}
where we have introduced the dimensionless parameters defined as $\tcs = \cs/(D_s\Lambda)$, $\tD = D_0/D_s$, and $\tl_i = \lambda_i/(\pi^{1/2}D_s\Lambda^{1/2})$. Eq.~(\ref{RG eq c lambda}) implies that the sound velocity and the nonlinear couplings are not renormalized.
To analyze the scaling behavior, we introduce the anomalous exponent $\anoexp_X$ for each parameter $X \in \{D_0, D_s, \cs, \lambda_i\}$, defined as $\anoexp_{X}=d(\log X)/d(\log \Lambda^{-1})$. This exponent describes the scaling relation $X \sim \Lambda^{-\anoexp_X}$. 
The RG equations for the dimensionless parameters are derived from their definitions,
using $\anoexp_{c_s}=\anoexp_{\lambda_i}=0$ obtained from Eq.~(\ref{RG eq c lambda}):
\begin{align}
  -\Lambda\frac{d \tcs}{d \Lambda} 
          &=\tcs\biggl(1-\anoexp_{D_s}\biggr), \label{eq:RG_tilde c}\\
     -\Lambda\frac{d  \tilde{D}}{d \Lambda} 
     &= \tilde{D}\biggl(\anoexp_{D_0}-\anoexp_{D_s}\biggr),\\
          -\Lambda\frac{d \tl_i}{d \Lambda} 
          &=\tl_i\biggl(\frac{1}{2}-\anoexp_{D_s}\biggl).\label{eq:RG_tilde lambda main}
\end{align}
These RG equations admit a nontrivial fixed point characterized by $z=3/2$, as detailed below.
The fixed points are obtained as zeros of the right-hand sides of Eqs.~(\ref{eq:RG_tilde c})-(\ref{eq:RG_tilde lambda main}) and the nontrivial solution reads
\begin{align}
    \tcs^*=0, \quad \anoexp_{D_0}^*=\anoexp_{D_s}^*=\frac{1}{2}.\label{eq:FP1 main}
\end{align}
Near this fixed point, we obtain the following anomalous scalings: $D_0 \sim \Lambda^{-1/2}$, $D_s \sim \Lambda^{-1/2}$, $\cs \sim \Lambda^{0}$, and $\lambda_i \sim \Lambda^0$. 

To determine the dynamical exponent $z$ of this fixed point, we consider the inverse Green function near the fixed point: $1/\Gf^\alpha(k, \omega) \sim \omega + i\bar{D}_\alpha \Lambda^{-1/2}k^2$, where $\bar{D}_\alpha$ is a constant independent of $\Lambda$. Note that the sound velocity vanishes at the fixed point. We will return to the scaling with finite $\cs$ later.
We perform the rescaling $k \to b^{-1}k$, $\Lambda \to b^{-1}\Lambda$, and $\omega \to b^{-z}\omega$, and obtain 
\begin{align}
       1/\Gf^\alpha( b^{-1}k, b^{-z}\omega) &= b^{-z} ( \omega + ib^{z-3/2}\bar{D}_\alpha \Lambda^{-1/2}k^2 ).
\end{align}
Then, the requirement of scaling invariance leads to $z=3/2$, which is the same as the KPZ exponent. Notably, both the sound and heat modes exhibit the same exponent, $z=3/2$. These exponents are the second result in this paper.
It should be noted that the exponent associated with the heat mode differs from the previously predicted value for the original NFH equations with the mode-coupling calculation~\cite{spohn2014nonlinear}.

\begin{figure*}[t]
\begin{center}
\includegraphics[scale=0.9]{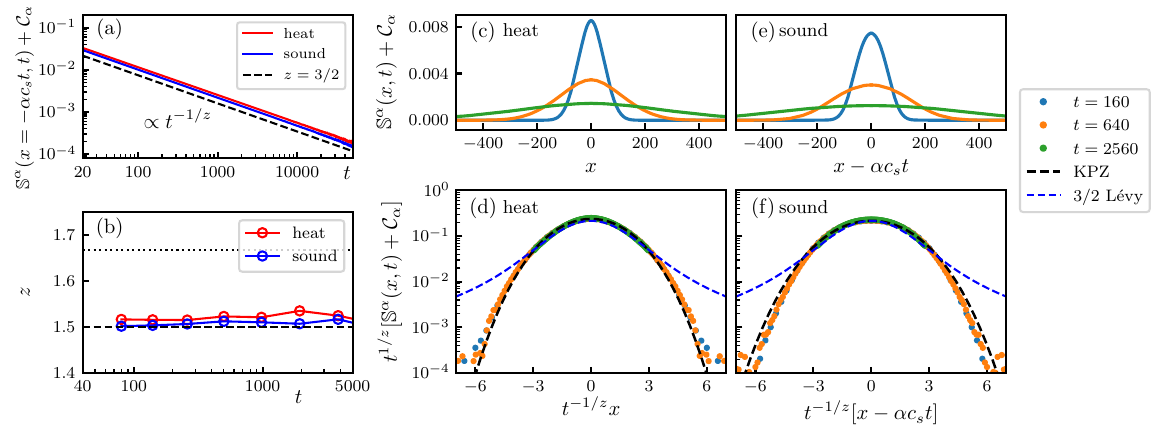}
\end{center}
\vspace{-0.5cm}
\caption{
Numerical results for the dynamical scaling of the space–time correlation $\Sf^\alpha(x,t)$ obtained from simulations of Eq.~(\ref{EOM phi main}) with parameters $D_0 = D_s = \lambda_{1,2,3,4} = 1$, $c_s = 0.1$ and system size $L=8192$. (a) Power-law decay of $\Sf^\alpha(x+\alpha c_s t=0,t)+\mathcal{C}_\alpha$ on a log–log scale in the comoving frame. (b) Dynamical exponent $z$ as a function of time $t$ extracted from the local slope in (a); the dashed line indicates the KPZ value $z=3/2$. (c), (e) Spatial profiles of $\Sf^\alpha(x,t)+\mathcal{C}_\alpha$ at various times, plotted against the comoving coordinate $x+\alpha c_s t$. (d), (f) Data collapse of (c) and (e) based on Eq.~(\ref{eq:modified_scaling}), revealing the universal scaling functions $f_\alpha$. 
The Pr\"{a}hofer–Spohn (black dashed) and $3/2$ L\'{e}vy (blue dashed) functions are shown for comparison. Additional numerical results are presented in Figs.~\ref{emfig1}–\ref{emfig3} in the End Matter.
}

\label{fig1}
\end{figure*}
\sectionprl{Numerical simulation}
To verify the RG predictions, we numerically integrate the equations for the normal modes~(\ref{EOM phi main}) 
using a staggered-grid discretization~\cite{BalboaUsabiaga2012-sh, Srivastava2023-nx, Garcia2024-nq} and a third-order stochastic Runge–Kutta scheme~\cite{Delong2013-fh, Srivastava2023-nx, Garcia2024-nq}. 
Details of the numerical implementation are provided in Sec.~VI of the SM~\cite{SM1}. 
This approach also enables us to analyze scaling functions beyond the reach of the RG framework. Notably, the exponent $z = 3/2$ does not by itself imply the universal KPZ scaling functions characteristic of the KPZ equation~\cite{roy2025fixed}.
To elucidate the scaling behavior, we compute the space–time correlation function 
$\Sf^\alpha(x,t) = \langle \phi^\alpha(x,t)\phi^\alpha(0,0)\rangle$, 
whose results are summarized in Fig.~\ref{fig1}.

Before discussing the results, we note that the numerical calculations are affected by finite-size effects due to the conservation law $\int dx\, \Sf^\alpha(x,t) = 0$ \footnote{Note that $\phi^{\alpha}=\sum_b R_{\alpha b } u_b$, where $u_b$ is the deviation from the average value, and hence $\int dx \phi^{\alpha} (x,t)=0$ holds identically}. To account for these effects in the scaling relations, we consider the modified scaling form~\cite{roy2025fixed} 
\begin{align}
\Sf^\alpha(x,t)  &= t^{-1/z} f_{\alpha}\!\left(\frac{x-\alpha c_s t}{t^{1/z}}\right) - \const_\alpha, \label{eq:modified_scaling}
\end{align} 
where $f_\alpha$ is a scaling function and $\const_\alpha$ is the offset arising from the finite-size effect. It is known that the finite-size effects in the fluctuating Navier-Stokes equation are well captured by the offset~\cite{Bell2007-yb}.
We have also numerically confirmed that the cross-correlations between the sound and heat modes, as well as between oppositely propagating sound modes, are negligibly small (see Fig.~\ref{emfig1} of the End Matter).
Below, we discuss the validity of the scaling relation Eq.~(\ref{eq:modified_scaling}).

We first examine the temporal correlation $\Sf^\alpha(x\!=\!0,t)$ to verify the predicted exponent $z=3/2$. 
From Eq.~(\ref{eq:modified_scaling}), it follows that $\Sf^\alpha(0,t) + \const_{\alpha} \sim t^{-1/z}$. 
The optimal offset $\const_\alpha$ is determined by fitting the data in the range $40 < t < 4000$ to $\mathcal{A}_\alpha t^{-1/z} - \const_\alpha$, with $z$, $\mathcal{A}_\alpha$, and $\const_\alpha$ as fitting parameters. 
Fig.~\ref{fig1}(a) shows a log–log plot of $\Sf^\alpha(0,t) + \const_{\alpha}$ using this optimized offset; 
the data for both modes align along straight lines consistent with the $t^{-2/3}$ scaling, i.e., $z=3/2$.

For a more quantitative analysis, we fit the local slope of the data in Fig.~\ref{fig1}(a) 
over intervals $(2+2^n)\!\times\!10 \le t < (2+2^{n+1})\!\times\!10$ ($n \in \mathbb{N}$). 
Fig.~\ref{fig1}(b) shows the resulting exponents $z$, which are in excellent agreement with the RG prediction $z=3/2$ for both the heat and sound modes.

Next, we consider whether the space–time correlations satisfy the dynamic scaling law in Eq.~(\ref{eq:modified_scaling}). We perform a data–collapse analysis using a neural–network regression~\cite{yoneda2023neural}, in which the scaling function $f_{\alpha}$ is represented by the network. The method fits the raw data $\Sf^\alpha(x,t)$ [Figs.~\ref{fig1}(c) and (e)] to Eq.~(\ref{eq:modified_scaling}), optimizing the exponent $z$, the offset $\const_{\alpha}$, and the neural–network parameters simultaneously. As shown in Figs.~\ref{fig1}(d) and (f), rescaling with the optimized $z$ and $\const_{\alpha}$ yields an excellent collapse of the data for both the heat and sound modes onto single universal curves, confirming the scaling law with finite-size corrections. 
The optimized exponents, $z=1.52$ for the heat mode and $z=1.51$ for the sound mode, are in excellent agreement with the RG predictions.

Finally, we discuss the explicit form of the scaling function $f_{\alpha}$. Although no theory provides its exact expression, previous studies~\cite{spohn2016fluctuating} suggest two prominent candidates: the Pr\"{a}hofer–Spohn scaling function, known as the exact solution of the KPZ equation, and the $3/2$ L\'{e}vy distribution. Figs~\ref{fig1}~(d) and (f) compare the numerically obtained universal curves with these theoretical forms. 
Both the heat and sound modes show clear proximity to the Pr\"{a}hofer–Spohn scaling function.
These results indicate that both modes follow KPZ universality at the level of two-point space–time correlations. Our numerical determination of these universal functions constitutes the third main result of this paper.

\sectionprl{Remarks on the sound velocities in the RG}
We now discuss the scaling of the inverse Green function with a finite sound velocity:
\begin{align}
       1/\Gf^\alpha&(b^{-1}k, b^{-3/2}\omega)\nonumber\\
       &= b^{-3/2}\biggl( \omega - b^{1/2}\alpha \cs k + i \bar{D}_\alpha \Lambda^{-1/2}k^2 \biggr). \label{scaling with cs}
\end{align}
Here, $\cs$ is assumed to be sufficiently small, and we are close to the fixed point (\ref{eq:FP1 main}). 
The sound velocity increases as $b^{1/2}$, and thus the scaling relation is apparently violated. 
This can be understood from the scaling behavior of the correlation function.
Numerical simulations show that the scaling relation between $x+\alpha \cs t$ and $t$ takes the form (\ref{eq:modified_scaling}).
If we rescale space and time in the standard RG manner, $x \to bx$ and $t \to b^z t$, we obtain
\begin{align}
    \Sf^\alpha(bx,b^z t)\sim b^{-1} t^{-1/z} f_\alpha \!\left(\frac{x-b^{z-1}\alpha \cs t}{t^{1/z}}\right).
\end{align}
The sound velocity increases as $b^{z-1}$, which is the same as Eq.~(\ref{scaling with cs}).
The difficulty is that the propagating mode obeys the scaling relation between $x-\cs t$ and $t$, instead of that between $x$ and $t$. 
The standard RG cannot directly describe this, and the relation can only be captured in the zero–sound–velocity limit. This happens even for the one-component Burgers equation with a finite velocity~\cite{SM2}. Furthermore, the coupled Burgers equation has three modes with different group velocities, so that these propagating velocities cannot be eliminated simultaneously by a coordinate transformation. 
Nevertheless, even in this case, numerical calculations show that each mode obeys the scaling form in Eq.~(\ref{eq:modified_scaling}). In particular, the simulations demonstrate that the correlation functions exhibit scaling even for finite sound velocity, and the exponent $z$ and the scaling functions are found to be independent of $\cs$ as shown in Fig.~\ref{emfig3} in the End Matter.

\sectionprl{Summary and concluding remarks}
We have formulated symmetry-based NFH equations, Eqs.~(\ref{EOM original main}) with (\ref{A matrix})–(\ref{H3 matrix main}) and Eqs.~(\ref{EOM phi main}) with (\ref{cs, D_0, D_s})–(\ref{parametrized Gmatrix}).
We then performed a dynamical RG analysis, identifying the fixed point with dynamical exponent $z=3/2$ for both sound and heat modes.
Extensive numerical simulations not only confirm this exponent but also reveal, from the space–time correlation functions, that both modes are well described by the universal KPZ scaling function, namely the Pr\"{a}hofer–Spohn function.
The RG analysis identifies the fixed-point structure, while the scaling law itself is supported by simulations. To our knowledge, this work is the first to formulate the RG issue associated with finite sound velocity, which standard RG cannot directly capture, and provides a first step toward clarifying this problem.

We also comment on systems with space-inversion symmetry 
$V(x)=V(-x)$, which imposes an additional constraint. Combined with time-reversal symmetry and the equilibrium condition, this constraint forbids nonlinear couplings at all orders, reducing the theory to a Gaussian one (see Sec.~II.B of the SM~\cite{SM1}) and yielding $z=2$. Without the equilibrium condition, the RG analysis instead exhibits unstable fixed points, possibly reflecting anomalous thermalization in the underlying Hamiltonian dynamics. A detailed study is left for future work.

This work provides a crucial step toward a deeper understanding of hydrodynamic behavior in Hamiltonian systems. We hope that our analysis can be extended to higher dimensions, such as two dimensions, which are relevant to graphene~\cite{Xu2014-oo}.

\section*{Acknowledgments}
We are deeply grateful to Herbert Spohn for his insightful and valuable comments.
YM is supported by JSPS KAKENHI Grant No. JP25K07148 and  the Ogawa science and technology foundation. HN is supported by JSPS KAKENHI Grant No. JP22K13978. KS is supported by JSPS KAKENHI Grant No. JP23K25796. The numerical computation in this study has been done using the facilities of the Supercomputer Center, the Institute for Solid State Physics, the University of Tokyo. 

\noindent
{\it Data availability.} - All data and codes that support the findings of this study are available on GitLab~\cite{DR} for public access.

\bibliography{refs}

\clearpage
\section*{End Matter}


\sectionprl{Appendix A: Explicit expressions for the symmetry-based NFH equations}
We present explicit expressions for the parameters in the symmetry-based NFH equations. See the SM for the details~\cite{SM1}. From the time-reversal symmetry, the inverse of the covariance matrix can be written in general as
\begin{align}
\Co^{-1} &= \begin{pmatrix}
a_1 & 0 & a_0 \\
0 & 1 & 0 \\
a_0 & 0 & a_3
\end{pmatrix}. 
\end{align}
We set $[\Co^{-1}]_{2,2}:=1$ without loss of generality. 
By requiring a Gaussian steady state of the Fokker–Planck equation, the following relations are obtained:
\begin{align}
D \Co =& B , \label{eq:FDT1 main}\\
A \Co  =& \Co A^T, \label{eq:FDT2}\\
\iCH^a_{bc} =&\iCH^b_{ca}=\iCH^c_{ab}, \label{potential condition main}
\end{align}
where $\iCH^a_{bc}:=\sum_d [\Co^{-1}]_{ad}H^d_{bc}$.
Combined with Eqs.~(\ref{eq:FDT1 main})–(\ref{potential condition main}), 
the conservation law $\partial_t u_1 - \partial_x u_2 = 0$ and 
time-reversal symmetry of the equations of motion 
yield the general expressions for the coefficients in Eq.~(\ref{EOM original main}):
\begin{align}
A &= \begin{pmatrix}
0 & -1 & 0 \\
a_0 P - a_1 & 0 & a_3 P - a_0 \\
0 & P & 0
\end{pmatrix}, \label{A matrix} \\
D 
&= \begin{pmatrix}
0 & 0 & 0 \\
0 & D_{22} & 0 \\
(a_0 / a_3) D_{33} & 0 & D_{33}
\end{pmatrix}, \\
B 
&= \begin{pmatrix}
0 & 0 & 0 \\
0 & D_{22} & 0 \\
0 & 0 & D_{33} / a_3
\end{pmatrix},\label{B matrix}\\
H^1 &= {\bm 0}, \\
H^2 &= \begin{pmatrix}
\Hgll& 0 & (a_3/a_0) \Hgll\\
0 & \Hggg & 0 \\
(a_3/a_0) \Hgll& 0 & (a_3^2 / a_0^2) \Hgll
\end{pmatrix},\\
H^3 &= \begin{pmatrix}
0 & (1/a_0) \Hgll & 0 \\
(1/a_0) \Hgll & 0 & (a_3/a_0^2) \Hgll \\
0 & (a_3/a_0^2) \Hgll & 0
\end{pmatrix}, \label{H3 matrix main}
\end{align}
where we have introduced the parameters $P=A_{32}$

By transforming the variables from $u$ to $\phi$, the equivalent general expressions for parameters automatically emerge as follows
\begin{align}
 \cs&=\sqrt{a_3 P^2-2 a_0 P+a_1}, \label{cs, D_0, D_s} \\
    D_0&=\frac{a_1 a_3-a_0^2 }{a_3 (a_3 P^2-2 a_0P+a_1)}D_{33}, \\
    D_s&=\frac{a_3^2 P^2-2 a_3 a_0 P+a_0^2 }{2 a_3 (a_3P^2-2 a_0 P+a_1)}D_{33}+\frac{1}{2}D_{22}, 
\end{align}
\begin{align}
    G^+&=\begin{pmatrix}
        -\lambda_3 & \lambda_2 & -\lambda_4\\
        \lambda_2 & \lambda_1 & 0 \\
        -\lambda_4 & 0 &\lambda_4
    \end{pmatrix}, \,
       G^-=\begin{pmatrix}
        -\lambda_4 & 0 & \lambda_4\\
        0 & -\lambda_1 & -\lambda_2 \\
        \lambda_4 &-\lambda_2& \lambda_3
    \end{pmatrix},\\
        G^0&=\begin{pmatrix}
        \lambda_2 & \lambda_1 & 0\\
         \lambda_1 & 0 & -\lambda_1 \\
        0 &-\lambda_1& -\lambda_2
    \end{pmatrix}. \label{parametrized Gmatrix}
\end{align}
Here, we parameterize the matrix $G$ with the parameters $\lambda_i~(i=1,2,3, 4)$.

\sectionprl{Appendix B: Renormalization group analysis}
Self energies by the one loop diagrams in Fig.~\ref{fig:self-energy} is explicitly written as
\begin{align}
\Xi^{\alpha}(k_1,\omega_1)=k_1^2&\sum_{\mu,\nu}G^\alpha_{\mu\nu}G^\alpha_{\mu\nu}\int^>_{k_2} \bSf^\mu_{k_+}\bSf^\nu_{k_-},\label{Xi main}\\
\Sigma^{\alpha}(k_1,\omega_1)=4k_1&\sum_{\mu,\nu} G^\alpha_{\mu\nu}G^\mu_{\alpha \nu}\int^>_{k_2} k_+ \Gf^\mu_{k_+}\bSf^\nu_{k_-},\label{Sigma main}\\
    \Gamma^\alpha_{\beta \gamma}(k_1,k_2, \omega_1,\omega_2)&= -4k_1 \sum_{\mu,\nu,\rho} G^\alpha_{\mu \rho}G^\beta_{\mu \nu}G^\gamma_{\nu\rho} {V}^\mu_{\nu\rho},\label{Gamma}
\end{align}
where $\Gf_k^{\alpha}:=1/(\omega -\alpha\cs k+i D_\alpha k^2)$ and $\bSf^\alpha_k:=2D_\alpha k^2/[(\omega-\alpha\cs k)^2+D_\alpha^2k^4]$. 
Similarly, vertex corrections given by the one loop diagram in Fig.~\ref{fig:vertex-corrections} yield the following expression
\begin{align}
V^{\mu}_{\nu\rho}&(k_1,k_2, \omega_1,\omega_2) =\int_q^> \bigl[
    q(k_++q) \Gf^\mu_{k_++q}\Gf^\nu_q\bSf^\rho_{k_--q}\nonumber \\ 
    &+q(k_-+q) \Gf^\rho_{k_-+q}\Gf^\nu_q\bSf^\mu_{k_+-q}
   \nonumber \\ 
   &+(k_++q)(k_--q)\Gf^\mu_{k_++q}\bSf^\nu_q\Gf^\rho_{k_--q}
    \bigr].   
\end{align}
These self energies and vertex corrections result in the RG equations~(\ref{RG eq Ds})--(\ref{RG eq c lambda}).
\usetikzlibrary{decorations.pathmorphing} 

\tikzset{
  every path/.style={line width=0.9pt, line cap=round, line join=round},
  extdot/.style={circle, fill, inner sep=1.8pt}, 
  wavy-clean/.style={
    decorate,
    decoration={snake, amplitude=1.8pt, segment length=6.2pt}
  }
}

\newcommand{\SelfEnergyA}{
\begin{tikzpicture}
  \coordinate (i) at (-1.6,0);
  \coordinate (a) at (0,0);
  \coordinate (b) at (6,0);
  \coordinate (f) at (7.6,0);

  \draw[shorten >=2pt] (i) -- (a);
  \draw[wavy-clean,shorten <=2pt] (b) -- (f);

  \draw[wavy-clean,shorten >=2pt,shorten <=2pt]
    (a) .. controls +(2.2,2.0) and +(-2.2,2.0) .. (b);
  \draw[shorten >=2pt,shorten <=2pt]
    (a) .. controls +(2.2,-2.0) and +(-2.2,-2.0) .. (b);

  \node[extdot] at (a) {};
  \node[extdot] at (b) {};

  \node[font=\Huge] at ($(a)!0.5!(b)+(0,0.15)$) {$\Sigma$};
\end{tikzpicture}%
}

\newcommand{\SelfEnergyB}{
\begin{tikzpicture}
  \coordinate (i) at (-1.6,0);
  \coordinate (a) at (0,0);
  \coordinate (b) at (6,0);
  \coordinate (f) at (7.6,0);

  \draw[wavy-clean,shorten >=2pt] (i) -- (a);
  \draw[wavy-clean,shorten <=2pt] (b) -- (f);

  \draw[shorten >=2pt,shorten <=2pt]
    (a) .. controls +(2.2,2.0) and +(-2.2,2.0) .. (b);
  \draw[wavy-clean,shorten >=2pt,shorten <=2pt]
    (a) .. controls +(2.2,-2.0) and +(-2.2,-2.0) .. (b);

  \node[extdot] at (a) {};
  \node[extdot] at (b) {};

  \node[font=\Huge] at ($(a)!0.5!(b)+(0,0.15)$) {$\Xi$};
\end{tikzpicture}%
}

\begin{figure}[htbp]
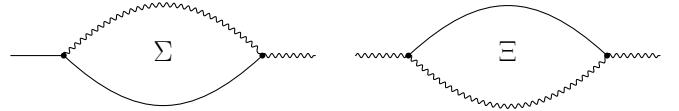

  \centering
  \resizebox{0.47\linewidth}{!}{\SelfEnergyA}\hfill
  \resizebox{0.47\linewidth}{!}{\SelfEnergyB}
  \caption{Self-energy diagrams. The left panel shows the $\Sigma$ loop corresponding to Eq.~(\ref{Sigma main}), and the right panel shows the  $\Xi$ loop corresponding to Eq.~(\ref{Xi main}).  The solid and wavy lines denote $\phi^\alpha$ and $\pi^\alpha$, respectively.}
  \label{fig:self-energy}
\end{figure}

\begin{figure}[htbp]
  \centering
  \resizebox{0.33\linewidth}{!}{\VCA}\hfill
  \resizebox{0.33\linewidth}{!}{\VCB}\hfill
  \resizebox{0.33\linewidth}{!}{\VCC}
  \caption{Diagrams of vertex corrections.}
  \label{fig:vertex-corrections}
\end{figure}

\begin{figure}[t]
\begin{center}
\includegraphics[scale=0.9]{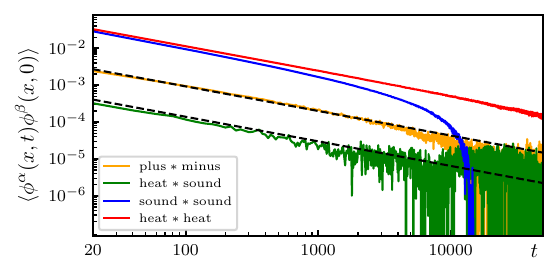}
\end{center}
\vspace{-0.5cm}
\caption{
Numerical results for the time correlation function $\langle \phi^{\alpha}(x,t)\phi^{\beta}(x,0)\rangle$. The parameters are identical to those in Fig.~\ref{fig1}. In contrast to Fig.~\ref{fig1}(a), which shows the autocorrelations ($\alpha=\beta$), this figure highlights the cross-correlations ($\alpha \neq \beta$), while also displaying the autocorrelations for reference. No constant offset $\mathcal{C}_{\alpha}$ is added, and the comoving frame ($x+\alpha c_s t=0$) is not used. The dashed line indicates a slope of $-2/3$, corresponding to the KPZ exponent.
}
\label{emfig1}
\end{figure}

\newpage

\sectionprl{Appendix C: Dynamical scaling behaviors of cross correlations}
This Appendix presents additional numerical data for the time correlation functions $\langle \phi^{\alpha}(x,t)\phi^{\beta}(x,0)\rangle$, supplementing the results shown in Fig.~\ref{fig1} of the main text. While the main text primarily focused on the autocorrelations ($\alpha=\beta$), here we examine the cross-correlations ($\alpha \neq \beta$) using the full nonlinear model, Eq.~(\ref{EOM phi main}).

The results are shown in Fig.~\ref{emfig1}, which reveal two key findings. First, the magnitudes of the cross-correlations are significantly smaller than those of the autocorrelations, typically by a factor of $10^{-1}$–$10^{-2}$. This provides a direct numerical justification for an approximation made in our RG analysis. In the main text, we neglected the renormalization corrections to the off-diagonal components of the inverse Green’s function. The data in Fig.~\ref{emfig1} confirm that these components are indeed weak, validating our RG treatment as a well-justified first approximation.

Second, Fig.~\ref{emfig1} shows that the cross-correlation functions themselves exhibit clear dynamical scaling. 
They follow a power-law decay, $\sim t^{-1/z}$, in agreement with the dynamical scaling hypothesis. 
Notably, the fitted dynamical exponent $z$ is remarkably close to the theoretical KPZ value of $1.5$, matching that obtained for the autocorrelations.

\sectionprl{Appendix D: Robustness of dynamical scaling for different sound velocities}

\begin{figure}[htbp]
\begin{center}
\includegraphics[scale=0.9]{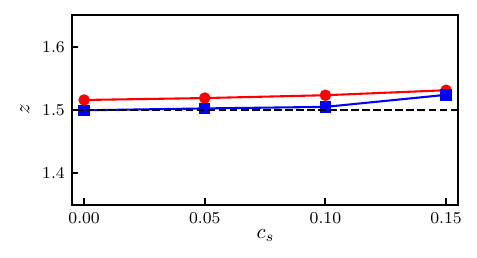}
\end{center}
\vspace{-0.5cm}
\caption{
Dependence of the dynamical critical exponent $z$ on the sound velocity $c_s$, where $z$ is obtained from the data collapse analysis of 
$\Sf^{\alpha}(x,t)$.
The error bars are comparable to the symbol size.
}
\label{emfig3}
\end{figure}

We present numerical evidence that the dynamical scaling law, Eq.~(\ref{eq:modified_scaling}), is robust against variations in the sound velocity $c_s$. 
The neural-network-based data-collapse analysis enables a high-precision determination of the dynamical exponent $z$. 
As summarized in Fig.~\ref{emfig3}, the fitted values of $z$ for all cases remain remarkably close to the theoretical KPZ value of $1.5$, showing little dependence on $c_s$. 
Additional results concerning the shape of the universal scaling function are provided in the SM~\cite{SM1}.

\clearpage
\onecolumngrid

\setcounter{secnumdepth}{2}

\renewcommand{\thesection}{\Roman{section}}
\setcounter{section}{0}

\renewcommand{\thesubsection}{\Alph{subsection}}
\setcounter{subsection}{0}

\setcounter{equation}{0}
\renewcommand{\theequation}{S\arabic{equation}}
\renewcommand{\thefigure}{S\arabic{figure}}
\renewcommand{\thetable}{S\arabic{table}}

\section*{Supplemental Material for\\
``Symmetry-based nonlinear fluctuating hydrodynamics in one dimension''}

\section{Fluctuating hydrodynamics and gradient expansion}
\label{FHD from derivative expansion}
In this section, we systematically construct the fluctuating hydrodynamics from the microscopic Hamiltonian. The construction proceeds by identifying conserved quantities, introducing the corresponding hydrodynamic variables, and writing down all terms allowed in a gradient expansion subject to the symmetry constraints, following the philosophy of effective field theory.

\subsection{Fluctuating current in the hydrodynamics}
We consider an $N$-particle system in one dimension, described by the Hamiltonian
\begin{align}
H &= \sum_{i} \frac{1}{2} p_i^2 +  \sum_{i } V(q_{i+1} - q_i), \label{supplhamil}
\end{align}
where $p_i$ and $q_i$ are the momentum and position of the $i$-th particle, respectively. Here, $i \in {\mathbb Z}/N {\mathbb Z}$, and we consider infinite line picture imposing the condition $p_{i+N}=p_i$ and $q_{i+N}= q_{i} + L$ where $L$ is the total length. The potential $V$ is an arbitrary interaction, where particles interact only with the nearest neighbor sites. 

The standard way to construct the hydrodynamics is to employ the coarse-graining in space. As in Ref. \cite{saito2021microscopic}, one may introduce the coarse-graining blocks where we assign the new coordinate $x$ containing $\ell_{\rm cg}$ numbers of particles. We focus only on the three conserved quantities out of totally $2N$ degrees of freedoms, i.e., local stretch arising from the conservation of the system size, local momentum, and local energy. This coarse-graining procedure is mathematically equivalent to project the microscopic Newtonian dynamics onto the three conserved quantities. This procedure is performed through Zwanzig's projection operator method (See, e.g., \cite{zwanzig,saito2021microscopic,zubarevmorozof}). Due to this projection, degrees of freedoms other than local conserved quantities play a role of thermal noise, where near equilibrium the fluctuation dissipation theorem should hold. 

Based on the above procedure, we discuss the general aspect of the local conserved quantities. Let $l(x,t)$, $g(x,t)$ and $e(x,t)$ be coarse-garined local stretch, momentum, and energy density. We now consider consider small fluctuations around thermal equilibrium,
\begin{align}
l(x,t) &= l_{\mathrm{eq}} + \delta l(x,t), \\
g(x,t) &= g_{\mathrm{eq}} + \delta g(x,t), \\
e(x,t) &= e_{\mathrm{eq}} + \delta e(x,t),
\end{align}
which we collectively denote by $\bm{u} = (\delta l, \delta g, \delta e)$. From the continuity equation, the local conserved quantities are connected to the local currents as
\begin{align}
\partial_t \bm{u} + \partial_x {\cal J} = \bm{0} \, ,
\end{align}
where ${\cal J} = ({\cal J}_l, {\cal J}_g, {\cal J}_e)$ is the corresponding currents. From the projection picture explained above, the local currents should have the fluctuating part in addition to the deterministic part as
\begin{align}
{\cal J} = {\cal J}^{\mathrm{det}} + {\cal J}^{\mathrm{ran}}\, ,
\end{align}
where the fluctuating part ${\cal J}^{\mathrm{ran}}$ is modeled as Gaussian white noise with correlation
\begin{align}
    \langle {\cal J}_a^\mathrm{ran}(x,t) {\cal J}_b^\mathrm{ran}(x',t')\rangle = 2B_{ab}\delta(t-t')\delta(x-x'), 
\end{align}
Here, $B$ is a symmetric matrix determined by symmetry.

\subsection{Gradient expansion}

The deterministic part ${\cal J}_{\mathrm{det}}$ is expanded in $\bm{u}$ to second order:
\begin{align}
    {\cal J}_a^\mathrm{det}&(x,t)=\sum_{b}\int dy\tilde{A}_{ab}(x-y) u_b(y,t)+\sum_{b,c}\int dydz \tilde{H}^a_{bc}(x-y,x-z)u_b(y,t)u_c(z,t)+O(u^3).
\end{align}
The expansion coefficients are generally nonlocal, but in the long-wavelength limit we expand $\tilde{A}$ and $\tilde{H}$ in spatial derivatives:
\begin{align}
\tilde{A}_{ab}(x-y) &\sim \biggl(A_{ab} - D_{ab} \partial_x + O(\partial_x^2)\biggr) \delta(x-y), \\
\tilde{H}^a_{bc}(x-y, x-z) &\sim \biggl(H^a_{bc} + O(\partial_x)\biggr) \delta(x-y) \delta(x-z).
\end{align}
Higher-order derivatives are irrelevant at long distances.
The resulting fluctuating hydrodynamic equation is
\begin{align}
\sum_{b} \biggl(\delta_{ab} \partial_t + A_{ab}  \partial_x - D_{ab}  \partial_x^2\biggr) u_b + \sum_{b,c} H^a_{bc}  \partial_x (u_b u_c)
+ \partial_x  {\cal J}_a^\mathrm{ran} = 0. \label{EOM original}
\end{align}
Note that the expansion is truncated at the relevant orders in the long-time and long-distance scaling limit, as discussed in Sec.~\ref{scaling analysis}.

\subsection{Equilibrium conditions}
\label{Derivation of equilibrium conditions}
We consider the isolated system described by the Hamiltonian (\ref{supplhamil}). Nevertheless, 
it is expected that the local conserved quantities exhibit thermalization, where the statistics of the local quantities is well described by the equilibrium distribution. The intensive variables such as effective temperature are determined through the initial condition.  
 
From this picture, we impose  a fundamental requirement associated with the thermal equilibrium. In our model, the equilibrium distribution is assumed to be
\begin{align}
\Pfp_{\mathrm{eq}}[u] &= \frac{1}{Z} \exp\left[-\frac{1}{2} \sum_{a,b} \int dx  u_a(x)  \Co_{ab}^{-1}  u_b(x)\right]\label{Peq}, \\
\Co^{-1} &= \begin{pmatrix}
a_1 & 0 & a_0 \\
0 & 1 & 0 \\
a_0 & 0 & a_3
\end{pmatrix}, \label{inverseC}
\end{align}
where the parameters $a_i$ are determined by the thermodynamic state, and the zeros arise from time-reversal symmetry. From the requirement that the Fokker–Planck equation associated with Eq.~(\ref{EOM original}) admits the distribution (\ref{Peq}) as a solution, we obtain the following conditions: 
\begin{align}
D \Co =& B, \label{eq:FDT1}\\
A \Co  =& \Co A^T, \label{eq:FDT2}\\
\iCH^a_{bc} =&\iCH^b_{ca}=\iCH^c_{ab}. \label{potential condition}
\end{align}
where we have introduced $\iCH^a_{bc}=\sum_d\Co^{-1}_{ad}H^d_{bc}$.

Below, we derive the equilibrium conditions (\ref{eq:FDT1})-(\ref{potential condition}). 
The Fokker-Planck equation corresponding to Eq.~(\ref{EOM original}) is given by
\begin{align}
    \partial_t \Pfp&+\sum_{a,b}\int dx \biggl(\partial_x\frac{\delta}{\delta u_a}\biggr) \biggl[A_{ab}u_b+\sum_cH^a_{bc}u_bu_c-D_{ab}\partial_xu_b-
    B_{ab} \biggl(\partial_x\frac{\delta}{\delta u_{b}}\biggr)\biggr]\Pfp=0,
\end{align}
where $\Pfp[\bm{u}]$ is the probability distribution function.
The detailed balance condition reads 
\begin{align}
\biggl[-D_{ab}\partial_xu_b-
B_{ab}
\biggl(\partial_x\frac{\delta}{\delta u_{b}}\biggr)\biggr]\Pfp_{\mathrm{eq}}=0.
\end{align}
When the equilibrium distribution is given by the form (\ref{Peq}), we obtain Eq.~(\ref{eq:FDT1}).

In addition to the detailed balance, the following condition remains for the Fokker–Planck equation to admit the equilibrium distribution as a solution:
\begin{align}
   \sum_{a,b}\int dx \biggl(\partial_x\frac{\delta}{\delta u_a}\biggr) 
   \biggl(A_{ab}u_b+\sum_cH^a_{bc}u_bu_c\biggr)\Pfp_{\mathrm{eq}}=0, 
\end{align}
for arbitrary $u_a(x)$.
This condition can be decomposed order by order in $u_a$, leading to the following relations
\begin{align}
  \sum_{a,b}\int dx \biggl(\partial_x\frac{\delta}{\delta u_a}\biggr) 
   A_{ab}u_b\Pfp_{\mathrm{eq}}&=-\sum_{a,b,c}\int dx 
   (\partial_xu_c)\Co^{-1}_{ca}A_{ab}u_b\Pfp_{\mathrm{eq}}=0,\label{A condition}\\
        \sum_{a,b,c}\int dx \biggl(\partial_x\frac{\delta}{\delta u_a}\biggr) H^a_{bc}u_bu_c\Pfp_{\mathrm{eq}}&= -\sum_{a,b,c,d}\int dx  (\partial_xu_d)\Co^{-1}_{da}H^a_{bc}u_bu_c\Pfp_{\mathrm{eq}}=0 \label{H condition}.
\end{align}

After dividing by $\Pfp_{\mathrm{eq}}$, we obtain
\begin{align}
\sum_{b,c}\int dx \, (\partial_x u_c) K_{cb} u_b &= 0, \label{A condition2} \\
\sum_{b,c,d}\int dx \, (\partial_x u_d) \iCH^d_{bc} u_b u_c &= 0, 
\label{H condition2}
\end{align}
where we have introduced the matrix $K := \Co^{-1} A$.

We first derive Eq.~(\ref{eq:FDT2}) from Eq.~(\ref{A condition2}).  
To this end, we decompose $K$ into its symmetric and antisymmetric parts,
\begin{align}
K = K^S + K^A,
\end{align}
where $(K^S)^T = K^S$ and $(K^A)^T = -K^A$.
For the symmetric part $K^S$, we obtain
\begin{align}
\sum_{b,c}\int dx \, (\partial_x u_c) K^S_{cb} u_b
&= \frac{1}{2} \sum_{b,c}\int dx \, \partial_x \!\left( u_c K^S_{cb} u_b \right) \\
&= \frac{1}{2} \sum_{b,c} \Bigl[ u_c K^S_{cb} u_b \Bigr]_{\text{boundary}} \\
&= 0 ,
\end{align}
where we have imposed periodic boundary conditions, under which the boundary contributions vanish.
Next, the antisymmetric part $K^A$ must satisfy
\begin{align}
\sum_{b,c}\int dx \, (\partial_x u_c) K^A_{cb} u_b = 0 .
\end{align}
Since this condition must hold for arbitrary periodic functions $u_a(x)$, it follows that $K^A = 0$. 
Therefore, we obtain
\begin{align}
\Co^{-1} A = (\Co^{-1} A)^T .
\end{align}
Noting that $\Co^T = \Co$, we arrive at Eq.~(\ref{eq:FDT2}).

To derive Eq.~(\ref{potential condition}), we introduce the functional
\begin{align}
F^a[u] := \sum_{b,c} \iCH^a_{bc} u_b u_c .
\end{align}
Using periodic boundary conditions, Eq.~(\ref{H condition2}) can be rewritten as
\begin{align}
\sum_a \int dx \, (\partial_x u_a) F^a[u]
&= \sum_a \oint du_a \, F^a[u] \\
&= 0 .
\end{align}
Since the integral over any closed path in $u$-space vanishes, the one-form $\sum_a F^a[u] \, du_a$ must be exact. Consequently, we obtain
\begin{align}
\frac{\delta F^a[u]}{\delta u_b}
=
\frac{\delta F^b[u]}{\delta u_a}.
\end{align}
Using this relation together with $\iCH^a_{bc} = \iCH^a_{cb}$, we arrive at Eq.~(\ref{potential condition}).

\section{Symmetry-based fluctuating hydrodynamics}
In this section, we derive the symmetry-based fluctuating hydrodynamics 
by imposing covariance on the hydrodynamic equation~(\ref{EOM original}) 
under a set of symmetries, which leads to constraints on the parameters of the equations. 
By combining these constraints with the equilibrium conditions~(\ref{eq:FDT1})–(\ref{potential condition}), 
we obtain the symmetry-based fluctuating hydrodynamics.

\subsection{General case}
\label{sec:generalcase}
Here, we determine the expansion coefficients $A$, $B$, $D$, and $H$ solely from conservation laws and symmetry considerations. The stretch variables are directly connected to the local momentum (see, e.g., \cite{saito2021microscopic}), i.e., 
\begin{align}
\partial_t u_1 - \partial_x u_2 &= 0 \, . \label{u1u2prop}
\end{align}
Hence, we find
\begin{align}
A_{1a} &= (0, -1, 0), \label{supple_a1}\\
{\cal J}_{\mathrm{ran}}^1(x,t)&=0,\\
D_{1a} &= B_{1a} = (0, 0, 0), \\
H^1_{ab} &= \mathcal{O}, \label{supple_h1}
\end{align}
where $\mathcal{O}$ denotes the zero matrix.

We next discuss time-reversal symmetry. The  $A$ and $D$ terms correspond to time-reversible and time-irreversible contributions, respectively. The $H$ term is time-reversible due to thermodynamic stability. 
Under time reversal transformation
\begin{align}
t \to -t, \quad u_1 \to u_1,\quad  u_2 \to -u_2,\quad u_3 \to u_3,   \label{tr_ope} 
\end{align}
reversible terms transform in the same way as the time-derivative term, while irreversible terms transform with the opposite sign. 
To satisfy time-reversal covariance, several matrix elements must vanish, and the coefficients can be generically written as
\begin{align}
A &= \begin{pmatrix}
0 & -1 & 0 \\
A_{21} & 0 & A_{23} \\
0 & A_{32} & 0
\end{pmatrix}, \label{A matrix 0} \\
D &= \begin{pmatrix}
0 & 0 & 0 \\
0 & D_{22} & 0 \\
D_{31} & 0 & D_{33}
\end{pmatrix}, \label{D matrix 0}\\
H^2 &= \begin{pmatrix}
H^2_{11} & 0 & H^2_{13} \\
0 & H^2_{22} & 0 \\
H^2_{13} & 0 & H^2_{33}
\end{pmatrix}, \\
H^3 &= \begin{pmatrix}
0 & H^3_{12} & 0 \\
H^3_{12} & 0 & H^3_{23} \\
0 & H^3_{23} & 0
\end{pmatrix}.
\end{align}
From Eqs.~(\ref{inverseC}), (\ref{eq:FDT1}) and (\ref{D matrix 0}), we obtain
\begin{align}
D &= \begin{pmatrix}
0 & 0 & 0 \\
0 & D_{22} & 0 \\
(a_0 / a_3) D_{33} & 0 & D_{33}
\end{pmatrix}, \label{D matrix} \\
B &=\begin{pmatrix}
0 & 0 & 0 \\
0 &  D_{22} & 0 \\
0 & 0 & D_{33} / a_3
\end{pmatrix}.\label{B matrix}
\end{align}
From  Eqs.~(\ref{inverseC}), (\ref{eq:FDT2}) and (\ref{A matrix 0}), we get the relations
\begin{align}
A_{21} = -a_1 + a_0 A_{32}, \quad A_{23} = -a_0 + a_3 A_{32}.
\end{align}
Defining $P = A_{32}$, we write
\begin{align}
A = \begin{pmatrix} 
0 & -1 & 0 \\
a_0 P - a_1 & 0 & a_3 P - a_0 \\
0 & P & 0
\end{pmatrix}.\label{A matirx}
\end{align}
Finally, the condition (\ref{potential condition}) provides the relations
\begin{align}
H^2_{11} &= a_0 H^3_{12},\label{pot cond1} \\
H^2_{33} &= a_3 H^3_{23}, \\
H^2_{13} &= a_3 H^3_{12}, \\
H^2_{13} &= a_0 H^3_{23}. \label{pot cond4}
\end{align}
These relations reduce the number of independent components of $H$ to two, namely $H^2_{11}$ and $H^2_{22}$. We thus obtain
\begin{align}
H^2 &= \begin{pmatrix}
\Hgll & 0 & (a_3/a_0) \Hgll \\
0 & \Hggg & 0 \\
(a_3/a_0) \Hgll & 0 & (a_3^2 / a_0^2) \Hgll
\end{pmatrix},\label{H2 matrix} \\
H^3 &= \begin{pmatrix}
0 & (1/a_0) \Hgll & 0 \\
(1/a_0) \Hgll & 0 & (a_3/a_0^2) \Hgll \\
0 & (a_3/a_0^2) \Hgll & 0
\end{pmatrix}. \label{H3 matrix}
\end{align}

\subsection{Special case with the space-inversion symmetry}
Here, we consider the special case where the Hamiltonian has the space-inversion symmetry, $i \to N-i$. This case is special, since this symmetry is rarely realized in realistic materials. In addition to Eqs.(\ref{eq:FDT1})-(\ref{potential condition}), (\ref{tr_ope}), and (\ref{supple_a1})-(\ref{supple_h1}), we need to take account of the covariance of the fluctuating hydrodynamics under the transformation 
\begin{align}
 x \to -x, \quad u_1 \to -u_1, \quad u_2 \to u_2,\quad  u_3 \to u_3.  \label{supp_invs}
\end{align}
This covariance imposes additional constraints, forbidding certain components of the coefficients. 
The coefficient matrices with space-inversion symmetry take the form
\begin{align}
\Co^{-1} &=  \begin{pmatrix} 
a_1 & 0 & 0 \\
0 & 1 & 0 \\
0 & 0 & a_3
\end{pmatrix}, \label{Cinverse matirx symm}\\
A &= \begin{pmatrix} 
0 & -1 & 0 \\
- a_1 & 0 & 0 \\
0 & 0 & 0
\end{pmatrix}, \label{A matirx symm}\\
D &= \begin{pmatrix}
0 & 0 & 0 \\
0 & D_{22} & 0 \\
0 & 0 & D_{33}
\end{pmatrix}, \label{D matrix symm}\\
B &=\begin{pmatrix}
0 & 0 & 0 \\
0 & D_{22} & 0 \\
0 & 0 & D_{33} / a_3
\end{pmatrix}, \label{B matrix symm}\\
H^1 &= H^2 = H^3 = \mathcal{O}.
\end{align}
We emphasize that all the $H$ matrices vanish under space-inversion symmetry. 

We show that any orders of nonlinear coefficients vanish under space-inversion symmetry. Consider the deterministic current of the form
\begin{align}
{\cal J}^{\rm det}_{a} &= \sum_b A_{ab} u_b + \sum_{b_1, \cdots , b_n} H^a_{b_1 , \cdots , b_n}  u_{b_1} \cdots u_{b_n} \, ,
\end{align}
where $H^a_{b_1 , \cdots , b_n}$ is the $n$th order nonlinear coefficient which is symmetric under any exchange between $b_i$ and $b_j$, i.e., $H^a_{b_1, \cdots, b_i \cdots, b_j , \cdots b_n}=H^a_{b_1, \cdots, b_j \cdots, b_i , \cdots b_n}$. In this case, through the same procedure as in Sec.\ref{Derivation of equilibrium conditions}, Eq.(\ref{potential condition}) in the equilibrium condition is replaced by 
\begin{align}
\hat{H}_{c,(b_1, \cdots , b_n)} &=\hat{H}_{b_j, (b_1,\cdots, c, \cdots, b_n)} \, , ~~~\forall j 
\label{ncyclic} \, , 
\end{align}
where 
\begin{align}
\hat{H}_{c,(b_1, \cdots , b_n)} &:=\sum_{a} ({\Co}^{-1})_{c a} H^{a}_{b_1, \cdots, b_n} \, . 
\end{align}
Note that the matrix ${\Co}^{-1}$ is diagonal as in (\ref{Cinverse matirx symm}) and hence $\hat{H}_{c,(b_1, \cdots , b_n)}$ is proportional to $H^{a}_{b_1, \cdots, b_n}$ as
\begin{align}
\hat{H}_{c,(b_1, \cdots, b_n)} &=\left\{ 
\begin{array} {ll}
0 & c=1 \, \\
H^{2}_{b_1, \cdots, b_n} & c=2 \, \\ 
a_3 H^{3}_{b_1, \cdots, b_n} & c=3 \, 
\end{array}
\right.  \, , \label{hhatb1bn}
\end{align}
where we use the property (\ref{u1u2prop}) for $c=1$. If $(b_1, \cdots , b_n)$ contains $1$, the condition (\ref{ncyclic}) leads to $H^{a}_{b_1, \cdots, b_n}=0$ due to the expression (\ref{hhatb1bn}). When $(b_1, \cdots , b_n)$ does not contain $1$, the symmetry (\ref{supp_invs}) requires $H^{a}_{b_1, \cdots, b_n}=-H^{a}_{b_1, \cdots, b_n}$, which implies $0$. Hence, we conclude that all higher order coefficients must vanish in the dynamics.

\section{fluctuating hydrodynamics in normal modes  for the general case and path integral representation}
\subsection{Fluctuating hydrodynamics in normal modes  for the general case }
\label{sec:action1}
We derive the fluctuating hydrodynamics for the normal modes for the general case satisfying the parameters in the section \ref{sec:generalcase}. 
The inverse Green function corresponding to Eq.~(\ref{EOM original}) in Fourier space is
\begin{align}
    \tilde{\Gf}^{-1}(k,\omega) =\omega\mathcal{I} -kA+ik^2D.
\end{align}
where $\mathcal{I}$ is the $3 \times 3$ identity matrix. We diagonalize $\tilde{\Gf}^{-1}$ by the $k$-dependent matrix $R(k)$:
\begin{align}
    \Gf^{-1}(k, \omega)&:=R(k) \tilde{\Gf}^{-1}( k, \omega )R^{-1}(k),\\
    &=\begin{pmatrix}
     \omega-\cs k+iD_sk^2 &0 &0 \\
        0& \omega+iD_0k^2 &0\\
        0 & 0 & \omega+\cs k+iD_sk^2
    \end{pmatrix}
    +O(k^3),
\end{align}
where $c_s$ is the sound velocity, and $D_0$ and $D_s$ are the diffusion constants for the heat and sound modes, respectively.
These coefficients are obtained as
\begin{align}
 \cs&=\sqrt{a_3 P^2-2 a_0 P+a_1},\\
    D_0&=\frac{a_1 a_3-a_0^2 }{a_3 (a_3 P^2-2 a_0P+a_1)}D_{33},\\
    D_s&=\frac{a_3^2 P^2-2 a_3 a_0 P+a_0^2 }{2 a_3 (a_3P^2-2 a_0 P+a_1)}D_{33}+\frac{1}{2}D_{22}.
\end{align}

The normal modes are defined by
\begin{align}
    \phi_\alpha(k)=\sum_a R_{\alpha a}(k)u_a(k),
\end{align}
with $\alpha = +, 0, -$ corresponding to the right-moving sound, heat, and left-moving sound modes, respectively. 
The equilibrium correlation matrix in the normal-mode basis is given by
\begin{align}
\Cn(k) := R(k) \Co R^T(-k),
\end{align}
and has diagonal components of order $O(k^0)$ and off-diagonal components of order $O(k^1)$. 
We neglect the off-diagonal components and use the arbitrariness of $R$ 
to normalize the diagonal components, so that $\Cn = I$.

The nonlinear couplings in the normal-mode basis are given by
\begin{align}
    G^\alpha_{\beta\gamma}:=\sum_{a,b,c}R_{\alpha a}(0)H^a_{bc}R^{-1}_{b\beta}(0)R^{-1}_{c\gamma}(0),
\end{align}
where we have retained only the lowest-order term in $k$ since higher-order terms are irrelevant. These couplings satisfy
\begin{align}
    G^\alpha_{\beta \gamma}&=G^\alpha_{\gamma \beta},\\
    G^\alpha_{\beta \gamma}&=-G^{-\alpha}_{-\beta -\gamma},\label{G symmetry1}\\
    G^\alpha_{\beta \gamma}&=G^\beta_{\gamma \alpha}.\label{G symmetry2}
\end{align}
Here, the symmetries in Eqs.~(\ref{G symmetry1}) and (\ref{G symmetry2}) correspond to the time-reversal symmetry and the equilibrium condition (\ref{potential condition}), respectively.
Since the detailed expressions of the parameters introduced earlier is not required for the RG calculation, we reparametrize the $G$ matrix as
\begin{align}
    G^+&=\begin{pmatrix}
        -\lambda_3 & \lambda_2 & -\lambda_4\\
        \lambda_2 & \lambda_1 & 0 \\
        -\lambda_4 & 0 &\lambda_4
    \end{pmatrix},\label{parametrized_Gmatrix1}\\
    G^0&=\begin{pmatrix}
        \lambda_2 & \lambda_1 & 0\\
         \lambda_1 & 0 & -\lambda_1 \\
        0 &-\lambda_1& -\lambda_2
    \end{pmatrix},\\
       G^-&=\begin{pmatrix}
        -\lambda_4 & 0 & \lambda_4\\
        0 & -\lambda_1 & -\lambda_2 \\
        \lambda_4 &-\lambda_2& \lambda_3
    \end{pmatrix}. \label{parametrized_Gmatrix3}
\end{align}
Note that the expressions (\ref{parametrized_Gmatrix1})-(\ref{parametrized_Gmatrix3}) are equivalent to (\ref{supple_h1}), (\ref{H2 matrix}) and (\ref{H3 matrix}).
With these couplings, the fluctuating hydrodynamic equations in the normal modes read
\begin{align}
    \biggl(\partial_t  +\alpha \cs \partial_x -D_\alpha \partial_x^2 \biggr)\phi^\alpha +\sum_{\beta, \gamma}G^\alpha_{\beta\gamma}\partial_x(\phi^\beta\phi^\gamma)&=\zeta^\alpha,\label{EOM normal}\\
    \langle \zeta^\alpha(x,t)\zeta^\beta(x',t')\rangle&=-2D_\alpha \delta_{\alpha\beta}\partial_x^2\delta(t-t')\delta(x-x'),
\end{align}
where $D_+ = D_- = D_s$ and we have imposed the fluctuation-dissipation relation to determine the noise intensity.

\subsection{Path integral representation}
\label{sec:action2}
We represent the stochastic dynamics using the Martin–Siggia–Rose–Janssen–de~Dominicis (MSRJD) path-integral formalism. The partition functional $Z$ is written as
\begin{align}
    Z&= \int \mathcal{D} \phi^\alpha\mathcal{D}\pi^\alpha e^{-I_0[\phi,\pi]+I_{int}[\phi,\pi]},\\
    I_0[\phi,\pi]&=\sum_\alpha\int_k \frac{1}{2}
    \begin{pmatrix}
        \phi^\alpha_{-k} &\pi^\alpha_{-k}
    \end{pmatrix}
    \begin{pmatrix}
        0 &  1/g^\alpha_{-k}  
 \\
  1/g^\alpha_{k} 
& 2D_\alpha k^2
    \end{pmatrix}
    \begin{pmatrix}
        \phi^\alpha_{k} \\
        \pi^\alpha_{k}
    \end{pmatrix},\label{eq:I0}\\
I_{\mathrm{int}}[\phi,\pi]&=-\sum_{\alpha,\beta,\gamma}\int_{k_1}\int_{k_2}k_1 G^\alpha_{\beta\gamma}
 \pi^\alpha_{-k_1} \phi^{\beta}_{k_+}\phi^{\gamma}_{k_-}.  \label{eq:Iint}
\end{align}
Here, $\omega$ denotes the frequency, $k$ the wavenumber, and $\pi_k^\alpha$ the auxiliary field. 
We have introduced the following abbreviated notations:
\begin{align}
    \phi^\alpha_k &= \phi^\alpha(k,\omega), \\
    \int_k &= \int_{-\Lambda}^{\Lambda} \frac{dk}{2\pi} 
              \int_{-\infty}^{\infty} \frac{d\omega}{2\pi}, \\
    k_\pm &= \frac{1}{2}k_1 \pm k_2, \\
    \omega_\pm &= \frac{1}{2}\omega_1 \pm \omega_2,
\end{align}
where $\Lambda$ is the wavenumber cutoff.

The zeroth-order correlation functions are calculated as
\begin{align}
\langle \phi^\alpha_{k_1}\pi_{k_2}^\beta  \rangle_0
 &=\frac{1}{Z_0}\int \mathcal{D}\phi^\alpha\mathcal{D}\pi^\alpha \phi^\alpha_{k_1}  \pi_{k_2}^\beta  e^{-I_0},\\
    &= g^\alpha_{k_1}
\delta_{\alpha\beta}\delta_{k_1,-k_2},\\
\langle \phi_{k_1}^\alpha \phi^\beta_{k_2} \rangle_0
    &= \bSf^\alpha_{k_1}\delta_{\alpha\beta}\delta_{k_1,-k_2},\\
    \langle \pi_{k_1}^\alpha \pi^\beta_{k_2} \rangle_0&=0,
\end{align}
where we have introduced 
\begin{align}
    Z_0&= \int \mathcal{D}\phi^\alpha\mathcal{D}\pi^\alpha   e^{-I_0},\\
     g_k^\alpha
&=\frac{1}{\omega -\alpha\cs k+iD_\alpha k^2},\\
    \bSf^\alpha_{k} &= \frac{2D_\alpha k^2}{(\omega-\alpha \cs k)^2+D_\alpha^2k^4},\\
    \delta_{k_1,-k_2}&=(2\pi)^2\delta(k_1+k_2)\delta(\omega_1+\omega_2).
\end{align}

\section{Relevance and irrelevance of terms from scaling analysis}
\label{scaling analysis}
We estimate the relevance or irrelevance of each term at long distances by performing a scaling analysis. 
For simplicity, we consider the following fluctuating hydrodynamic equation for a single-component field $\phi(x,t)$:
\begin{align}
    (\partial_t - D \partial_x^2)\phi 
    + \sum_{n=1}^\infty \sum_{m=2}^\infty 
      \lambda_{n,m} \, \partial_x^n \phi^m 
    &= \zeta , \\
    \langle \zeta(x,t) \zeta(x',t') \rangle 
    &= -2 D \partial_x^2 
       \delta(x - x') \delta(t - t') ,
\end{align}
where this expression represents the most general nonlinear contribution consistent with a derivative expansion in $\partial_x$ and $\phi$, and $\lambda_{n,m}$ denotes the corresponding expansion coefficients.

We assume that the noise $\zeta$ satisfies the fluctuation–dissipation relation and that $\phi$ is a conserved field, implying $n \geq 1$.
This form corresponds to the expansion of ${\cal J}$ discussed in Sec.~\ref{FHD from derivative expansion}. 
Note that the number of components of $\phi$ does not affect the following argument.
The corresponding MSRJD action is given by
\begin{align}
 I_0 &= \int dt\, dx \, \bigl[i\pi(\partial_t - D\partial_x^2)\phi - \pi D\partial_x^2\pi\bigr],\\
    I_{\mathrm{int}} &= \int dt\, dx \sum_{n=1}^\infty \sum_{m=2}^\infty \lambda_{n,m}\,i\pi\,\partial_x^n\phi^m.
\end{align}

We next perform the scaling transformation
\begin{align}
    x \to bx, \quad 
    t \to b^z t, \quad 
    \phi \to b^{\alpha_{\phi}}\phi, \quad 
    \pi \to b^{\alpha_{\pi}}\pi, 
    \label{scaling transformation}
\end{align}
where $z$, $\alpha_\phi$, and $\alpha_\pi$ are scaling exponents determined from scale invariance.
Under this transformation, the Gaussian part of the action, $I_0$, becomes
\begin{align}
    I_0 = \int dt\, dx \, \biggl[
        b^{\alpha_\phi+\alpha_\pi+1}i\pi\bigl(\partial_t - b^{z-2}D\partial_x^2\bigr)\phi
        - b^{z+2\alpha_\pi-1}D\pi \partial_x^2\pi
    \biggr].
\end{align}
The action $I_0$ remains invariant under the choice
\begin{align}
    z = 2, \quad \alpha_\phi = \alpha_\pi = -\tfrac{1}{2}, \label{EW exponent}
\end{align}
which corresponds to the Edwards–Wilkinson (EW) fixed point.
For the EW exponents, the interaction term becomes
\begin{align}
    I_{\mathrm{int}}=\int dt\, dx \sum_{n,m} b^{5/2-m/2-n}\,\lambda_{n,m}\,i\pi\,\partial_x^n \phi^m.
\end{align}
Terms with  positive (negative) scaling exponents are relevant (irrelevant), while the choice $b>1$ corresponds to probing the long-distance behavior. 
For $n \ge 1$ and $m \ge 2$, the only relevant combination is
\begin{align}
    n = 1, \quad m = 2. \label{sup:relevant}
\end{align}
The term with $n = 1$ and $m = 3$ is marginal, having zero scaling exponent, 
and may contribute to anomalous scaling in the RG analysis. 
However, we can neglect it, since the relevant term~(\ref{sup:relevant}) 
dominates the scaling behavior. 
We also remark that the derivation of the model in Sec.~\ref{FHD from derivative expansion} 
can be made more systematic within the MSRJD path-integral formalism, 
which provides a natural framework for constructing an effective field theory.

We also consider the Gaussian action with a finite sound velocity,
\begin{align}
    I_0&=\int dt\, dx \biggl[i\pi\bigl(\partial_t+\cs\partial_x-D\partial_x^2\bigr)\phi-D\pi \partial_x^2\pi\biggr].
\end{align}
Under the scaling transformation (\ref{scaling transformation}), $I_0$ transforms as
\begin{align}
    I_0=\int dt\, dx \biggl[b^{\alpha_\phi+\alpha_\pi+1}i\pi\bigl(\partial_t+b^{z-1}\cs\partial_x-b^{z-2}D\partial_x^2\bigr)\phi
    -b^{z+2\alpha_\pi-1}D\pi \partial_x^2\pi\biggr].
\end{align}
Choosing $z=1$ renders the diffusion constant irrelevant,
\begin{align}
    D \to b^{-1}D,
\end{align}
which indicates that the theory becomes nondissipative at large scales and is therefore unphysical.
For the EW exponents (\ref{EW exponent}), we find
\begin{align}
    I_0=\int dt\, dx \biggl[i\pi\bigl(\partial_t+b \cs\partial_x-D\partial_x^2\bigr)\phi-D\pi \partial_x^2\pi\biggr],
\end{align}
where $\cs$ appears to increase with $b$, thereby breaking scale invariance. 
We argue, however, that this apparent scaling of $\cs$ is superficial rather than physical. 
A more detailed discussion of this issue will be given in the main text and in Sec.~\ref{about sound speed}.

\section{Renormalization group analysis}
In Wilson’s renormalization group, we integrate out modes in the wavenumber shell $\Lambda - \delta \Lambda \leq |k| \leq \Lambda$. 
We decompose the fields into short- and long-wavelength components:
\begin{align}
    \phi^\alpha_k &=\phi^{\alpha,>}_k+\phi^{\alpha,<}_k,\\
    \phi^{\alpha,>}_k &=\theta(\Lambda-\delta\Lambda-|k|)\phi^\alpha_k,\\
    \phi^{\alpha,<}_k &=\theta(|k|-\Lambda+\delta\Lambda)\phi^\alpha_k,
\end{align}
where  $\theta(k)$ is the Heaviside step function.
$\pi^\alpha$ is also decomposed in the same way.
Integrating over the short-wavelength components, we obtain a coarse-grained effective action $I_R$:
\begin{align}
    Z&=\int\mathcal{D}\phi^{\alpha,<} \mathcal{D}\pi^{\alpha,<} \mathcal{D}\phi^{\alpha,>} \mathcal{D}\pi^{\alpha,>} e^{-I_0+I_{\rm int}},\\
    &=\int\mathcal{D}\phi^{\alpha,<} \mathcal{D}\pi^{\alpha,<} e^{-I_R}.
\end{align}
which is formally expressed in terms of the self-energies $\Sigma^\alpha(k,\omega)$ and $\Xi^\alpha(k,\omega)$, and vertex corrections $\Gamma^\alpha_{\beta\gamma}(k_1,k_2,\omega_1,\omega_2)$:
\begin{align}
    I_R
    =&\int_k \frac{1}{2}
    \begin{pmatrix}
        \phi^{\alpha,<}_{-k} &\pi^{\alpha,<}_{-k}
    \end{pmatrix}
    \begin{pmatrix}
        0 & 1/g^\alpha_{-k} 
+\Sigma_{-k}^\alpha \\
    1/g^\alpha_{k}
+\Sigma_{k}^\alpha  & 2D_\alpha k^2+2\Xi^\alpha_k
    \end{pmatrix}
    \begin{pmatrix}
        \phi^{\alpha,<}_{k} \\
        \pi^{\alpha,<}_{k}
    \end{pmatrix}\nonumber \\
    &-\int_{k_1}\int_{k_2}\biggl[\biggl(k_1 G^\alpha_{\beta\gamma}+\Gamma^\alpha_{\beta\gamma}\biggr)\pi^{\alpha,<}_{-k_1}\phi^{\beta,<}_{k_+}\phi^{\gamma,<}_{k_-} \biggr],
\end{align}

In the Wilson RG,  the self-energies and vertex corrections are computed perturbatively in $G^\alpha_{\beta\gamma}$.
From the one-loop calculation, we obtain the following self energies (See the diagrams shown in Figs.2 and 3 of the End matter in the main text.) 
\begin{align}
\Xi^{\alpha}(k,\omega)=k^2\sum_{\mu,\nu}G^\alpha_{\mu\nu}G^\alpha_{\mu\nu}\int^>_{k_2} \bSf^\mu_{k_+}\bSf^\nu_{k_-},\label{Xi}\\
\Sigma^{\alpha}(k,\omega)=4k\sum_{\mu,\nu} G^\alpha_{\mu\nu}G^\mu_{\alpha \nu}\int^>_{k_2} k_+ g^\mu_{k_+}
\bSf^\nu_{k_-},\label{Sigma}
\end{align}
where we have introduced the notations
\begin{align}
    \int_{k}^>&=\int_{\Lambda-\delta\Lambda \leq |k|\leq \Lambda}\frac{dk}{2\pi}\int^{\infty}_{-\infty}\frac{d\omega}{2\pi},
\end{align}
They satisfy the relation $\operatorname{Im} \Sigma^\alpha_k = \Xi^\alpha_k$ which corresponds to the fluctuation–dissipation theorem. 
The vertex correction at one-loop order is calculated as 
\begin{align}
    \Gamma^\alpha_{\beta \gamma}(k_1,k_2,\omega_1,\omega_2)=& -4k_1 \sum_{\mu,\nu,\rho} G^\alpha_{\mu \rho}G^\beta_{\mu \nu}G^\gamma_{\nu\rho} {V}^\mu_{\nu\rho}(k_1,k_2,\omega_1,\omega_2),\label{Gamma}\\
    V^{\mu}_{\nu\rho}(k_1,k_2,\omega_1,\omega_2):=&\int_q^> \biggl[
    q(k_++q)  g^\mu_{k_++q} g^\nu_q
\bSf^\rho_{k_--q} +q(k_-+q) g^\rho_{k_-+q} g^\nu_q
\bSf^\mu_{k_+-q} +(k_++q)(k_--q)g^\mu_{k_++q}\bSf^\nu_q g^\rho_{k_--q}
    \biggr].\label{eq:def of V}
\end{align}

We define the renormalization corrections to the parameters as the expansion coefficients of the self energies and vertex correction in $k$ and $\omega$
\begin{align}
    \delta D_0 &=\frac{1}{2i}\lim_{\omega, k \to 0}\partial_k^2\Sigma^0_k, \\
    \delta D_s &=\frac{1}{2i}\lim_{\omega, k \to 0}\partial_k^2\Sigma^\sigma_k, \\
    \delta \cs   &=-\frac{1}{\sigma}\lim_{\omega, k \to 0}\partial_k\Sigma^\sigma_k,\\
    \delta G^\alpha_{\beta\gamma} &=\lim_{\omega_1,\omega_2, k_1,k_2 \to 0}\partial_{k_1}\Gamma^\alpha_{\beta\gamma},
\end{align}
where $\sigma=\pm$.
The renormalization corrections to $\delta \lambda_i$ are given by $\delta G^\alpha_{\beta\gamma}$, with the indices corresponding to those in Eqs.~(\ref{parametrized_Gmatrix1})-(\ref{parametrized_Gmatrix3}):
\begin{align}
    \delta \lambda_1=\delta G^+_{00},\\
    \delta \lambda_2=\delta G^+_{+0},\\
    \delta \lambda_3=\delta G^-_{--},\\
    \delta \lambda_4=\delta G^+_{--}.
\end{align}

From Eqs.~(\ref{Xi})-(\ref{Gamma}), we obtain 
\begin{align}  
\delta D_s=&\delta\Lambda\biggl[\frac{\lambda _1^2}{  \pi \Lambda^2   D_0}+\frac{4 \lambda_2^2(D_0+D_s)}{\pi (c_s^2+\Lambda (D_0+D_s))^2}+\frac{\lambda _3^2}{  \pi \Lambda^2  D_s}+\frac{\lambda_4^2(\cs^2+3 \Lambda^2  D_s^2)}{    \pi \Lambda^2  D_s(   c_s^2+ \Lambda ^2 D_s^2)}\biggr],\\
\delta D_0 =& \delta \Lambda\biggl[\frac{8\lambda_1^2 (D_0+D_s)}{\pi  (c_s^2+\Lambda^2(D_0+D_s)^2)}+\frac{2\lambda_2^2}{\pi  \Lambda^2 D_s}\biggr], \\
\delta \cs =& \delta \lambda_i = 0.
\end{align}
where the subscripts $i=1,2,3$ and $4$. Note that the parameters $\lambda_i$ are defined in (\ref{parametrized_Gmatrix1})-(\ref{parametrized_Gmatrix3}).
Here, we note that the sound velocity and all the nonlinear couplings $\lambda_i$ do not have the RG corrections. 
In the one component noisy Burgers equation, a nonlinear coupling $\lambda$ is not renormalized due to the pseudo-Galilean symmetry.
Our multi-component model does not have  such simple Galilean invariance. Thus, the RG correction is written as
\begin{align}
    \delta G^\alpha_{\beta\gamma}=-4 \sum_{\mu,\nu,\rho} G^\alpha_{\mu \rho}G^\beta_{\mu \nu}G^\gamma_{\nu\rho} {V}^\mu_{\nu\rho}(0,0),
\end{align}
 where ${V}^\mu_{\nu\rho}(0,0)$ vanishes for all components $\mu$, $\nu$ and $\rho$.
This might indicate the existence of some extended pseudo-Galilean symmetry.

Dividing the RG corrections by $\delta \Lambda$, we obtain the RG equation  
\begin{align}
            -\Lambda\frac{d D_s}{d \Lambda} =&D_s \biggl[\frac{\lambda _1^2}{  \pi \Lambda   D_sD_0}+\frac{4 \lambda_2^2\Lambda(D_0+D_s)}{\pi D_s(c_s^2+\Lambda (D_0+D_s))^2}+\frac{\lambda _3^2}{  \pi \Lambda  D_s^2}+\frac{\lambda_4^2(\cs^2+3 \Lambda^2  D_s^2)}{    \pi \Lambda  D_s^2(   c_s^2+  \Lambda ^2 D_s^2)}\biggr],\\
    -\Lambda\frac{d D_0}{d \Lambda} =&D_0 \biggl[\frac{8\lambda_1^2 \Lambda(D_0+D_s)}{\pi D_0 (c_s^2+\Lambda^2(D_0+D_s)^2)}+\frac{2\lambda_2^2}{\pi  \Lambda D_0D_s}\biggr],\\
              -\Lambda\frac{d \cs}{d \Lambda} =&0,\label{eq:RG_c}\\
           -\Lambda\frac{d \lambda_i}{d \Lambda} =&0.\label{eq:RG_lambda}
\end{align}
For later convenience, we collectively denote the parameters $D_0, D_s, \cs$ and $\lambda_i$ by $X$, and introduce the corresponding function $\eta_X$ as in the following equation
\begin{align}
    -\Lambda \frac{d X}{d\Lambda}=X \cdot \eta_{X}.
\end{align}
We note that $\eta_X$ gives the anomalous exponent
\begin{align}
    X \sim \Lambda^{-\eta_X}.
\end{align}

\subsection{Fixed points and scaling exponents}
Fixed points of RG flow are defined in dimensionless parameter space. 
To define dimensionless parameters, we perform the dimensional analysis. 
From the fact that the dimension of the action $[I]$ is $1$, we obtain
\begin{align}
    [\omega]&=[D_\alpha\Lambda^2],\\
    [\pi^\alpha_k]&=[D_\alpha^{-1}\Lambda^{-5/2}],\\
    [\phi^\alpha_k]&=[D_\alpha^{-1}\Lambda^{-5/2}],\\
    [\cs]&=[D_\alpha \Lambda],\\
    [\lambda_i]&=[\Lambda^{1/2}D_\alpha].
\end{align}
 We make the parameters dimensionless on the basis of $D_s$ and $\Lambda$ as
\begin{align}
    \tcs &= \frac{\cs}{D_s\Lambda},\label{eq:Def of tilde c}\\
    \tD &= \frac{D_0}{D_s}, \label{eq:Def of tilde D}\\
    \tl_i &= \frac{\lambda_i}{\pi^{1/2}D_s\Lambda^{1/2}}, \label{eq:Def of tilde lambda}
\end{align}
In terms of the dimensionless parameters, the anomalous exponents are expressed as
\begin{align}
    \eta_{D_s}&=\frac{\tl_1^2}{\tD}+\frac{4 \tl_2^2(\tD+1) }{\tcs^2+(\tD+1)^2}+\tl_3^2+\frac{\tl_4^2(\tcs^2+3)}{\tcs^2+1}, \label{fdslambda}\\
    \eta_{D_0}&=\frac{2}{\tD}\biggl(\frac{4 (\tD+1) \tl_1^2}{\tcs^2+(\tD+1)^2}+\tl_2^2\biggr), \label{fd0lambda}\\
    \eta_{\cs}&=\eta_{\lambda_i}=0. \label{f c lambda}
\end{align}
Differentiating Eqs.~(\ref{eq:Def of tilde c})-(\ref{eq:Def of tilde lambda}) with respect to $\Lambda$ and using Eq. (\ref{f c lambda}),
 we arrive at the RG equations for the dimensionless parameters
\begin{align}
  -\Lambda\frac{d \tcs}{d \Lambda} 
          &=\tcs\biggl(1-\eta_{D_s}\biggr), \label{eq:eq:RG_tilde c}\\
     -\Lambda\frac{d  \tilde{D}}{d \Lambda} 
     &= \tilde{D}\biggl(\eta_{D_0}-\eta_{D_s}\biggr),\label{eq:RG_tildeddd}\\
          -\Lambda\frac{d \tl_i}{d \Lambda} 
          &=\tl_i\biggl(\frac{1}{2}-\eta_{D_s}\biggl),\label{eq:RG_tilde lambda}
\end{align}
The fixed points are given by zeros of the r.h.s of these equations. We have the two solutions 
\begin{align}
    \tcs^*=\tl_i^*=0,\label{eq:FP2}
\end{align}
and 
\begin{align}
    \tcs^*=0, \quad \eta_{D_0}^*=\eta_{D_s}^*=\frac{1}{2}.\label{eq:FP1}
\end{align}
The first solution (\ref{eq:FP2}) is the EW fixed point and trivial.
In contrast, the anomalous exponents (\ref{eq:FP1}) provide the non-trivial scalings of the parameters
\begin{align}
 D_0 \sim \Lambda^{-1/2},\; D_s \sim \Lambda^{-1/2},  \; 
&\cs \sim \Lambda^{0}, \; \lambda_i \sim \Lambda^0. 
\end{align}
By using Eqs.~(\ref{fdslambda}) and (\ref{fd0lambda}) with $\tilde{c}_s=0$, we find that the anomalous exponents $\eta_{D_0}=1/2$ and $\eta_{D_s}=1/2$ are realized with real parameters $\tilde{\lambda}_{i}~ (i=1,\cdots ,4)$ for the following two cases
\begin{align}
({\rm A}): 1 \le \tilde{D} \le 7 \, ,~~~~~~~~~~~~~({\rm B}): 0 < \tilde{D} \le 7 - 4 \sqrt{3} \, . 
\end{align}
For case (A), the parameters can be chosen as
\begin{align}
\tilde{\lambda}_2^2 &\le \frac{1}{4}\,
\frac{\tilde{D}(7-\tilde{D})(\tilde{D}+1)}{16 \tilde{D} - (\tilde{D}+1)^2},\\
\tilde{\lambda}_1^2 &= \frac{\tilde{D}(\tilde{D}+1)}{16}
 - \frac{(\tilde{D}+1)\tilde{\lambda}_2^2}{4},\\
\tilde{\lambda}_3^2 + 3 \tilde{\lambda}_4^2
 &= \frac{1}{2} - \frac{\tilde{\lambda}_1^2}{\tilde{D}}
 - \frac{4 \tilde{\lambda}_2^2}{\tilde{D}+1} \ (\ge 0).
\end{align}
Similarly, for case (B), the parameters are chosen as
\begin{align}
\frac{\tilde{D}}{4} &\ge \tilde{\lambda}_2^2 \ge
\frac{1}{4}\,\frac{\tilde{D}(7-\tilde{D})(\tilde{D}+1)}{16 \tilde{D} - (\tilde{D}+1)^2},\\
\tilde{\lambda}_1^2 &= \frac{\tilde{D}(\tilde{D}+1)}{16}
 - \frac{(\tilde{D}+1)\tilde{\lambda}_2^2}{4},\\
\tilde{\lambda}_3^2 + 3 \tilde{\lambda}_4^2
 &= \frac{1}{2} - \frac{\tilde{\lambda}_1^2}{\tilde{D}}
 - \frac{4 \tilde{\lambda}_2^2}{\tilde{D}+1} \ (\ge 0).
\end{align}
For both cases, we have numerically investigated the RG flow of Eqs.~(\ref{eq:RG_tildeddd}) and (\ref{eq:RG_tilde lambda}) with $\tilde{c}_s = 0$. 
We have confirmed that an attractive parameter regime exists. 
For the flow of the sound velocity, see the main text (Section~``Remarks on the sound velocities in the RG'').

To obtain the dynamic scaling exponent $z$, we consider the inverse of the Green function near the fixed point (\ref{eq:FP1})
\begin{align}
    (\Gf^{-1})^\alpha(\omega, k)&\sim \omega+i\bar{D}_\alpha  \Lambda^{-1/2}k^2,
\end{align}
where $\bar{D}_\alpha$ is the constant being independent of $\Lambda$. We note that the sound velocity vanishes at the fixed point, $\tcs^*=0$.
We perform the rescaling $k \to b^{-1}k$, $\Lambda \to b^{-1}\Lambda$ and $\omega \to b^{-z}\omega$ and have
\begin{align}
       (\Gf^{-1})^\alpha(b^{-z}\omega, b^{-1}k)&=b^{-z}\biggl( \omega+ib^{z-3/2}\bar{D}_\alpha  \Lambda^{-1/2}k^2\biggr).
\end{align}
Therefore, $\Gf^{-1}$ scales with $z=3/2$, which is the same value of the KPZ exponent.
We note that both of the heat and sound modes has the exponents $z=3/2$.

\subsection{  sound velocity and scaling relations}
\label{about sound speed}
We consider the one-component Burgers equation with a sound velocity
\begin{align}
\partial_t \phi+\partial_x\biggl(\cs \phi-\partial_x D \phi+ \lambda \phi^2
\biggr)&=\eta,\\
\langle\eta(x,t)\eta(x',t')\rangle &= -2D \partial_x^2\delta(x-x')\delta(t-t'), 
\end{align}
which is derived from the usual Burgers equation by the transformation $x \to x+\cs t$.
Then, the space-time correlation function exactly obeys Prah\"{o}fer–Spohn scaling function with the argument
\begin{align}
\Sf(x,t)= t^{-{2/3}}f \biggl(\frac{x-\cs t}{t^{2/3}}\biggr).
\end{align}
On the one hand, the one-loop calculation gives the following RG equations
\begin{align}
    -\Lambda\frac{\partial \cs}{\partial \Lambda}&=0,\\
        -\Lambda\frac{\partial \lambda}{\partial \Lambda}&=0,\\
    -\Lambda\frac{\partial D}{\partial \Lambda}&=D\tl^2,\label{eq:dD}\\
     -\Lambda\frac{\partial \tilde{c}_s}{\partial \Lambda}&=\tilde{c}_s\biggl(1-\tl^2\biggr),\\
      -\Lambda\frac{\partial \tl}{\partial \Lambda}&=\tl\biggl(\frac{1}{2}-\tl^2\biggr).
\end{align}
The KPZ fixed point reads
\begin{align}
    \tilde{c}_s^*=0, \quad (\tl^*)^2=\frac{1}{2},
\end{align}
where we again find that the sound velocity vanishes at the KPZ fixed point. 
Therefore, in the standard RG analysis, the dimensionless sound velocity apparently increases near the fixed point as
\begin{align}
    \tcs \sim \Lambda^{-1/2}.
\end{align}
This apparent increase is superficial, and the scaling exponent is independent of the sound velocity since it can be eliminated by the coordinate transformation.
The difficulty arises because the propagating mode obeys a scaling relation involving $x+\cs t$ and $t$, rather than $x$ and $t$. 
Consequently, the conventional RG scheme cannot straightforwardly capture this behavior and effectively treats the system in the limit of vanishing sound velocity.

\section{Details of numerical simulation}
In this appendix, we provide the details of the numerical simulation presented in the main text.
For the reader's convenience, we restate the equations to be solved:
\begin{align}
    \frac{\partial \phi^{\alpha}}{\partial t} + \frac{\partial \mathcal{J}^{\alpha}}{\partial x} = 0 \qquad \alpha=\pm,0
\end{align}
where the flux $\mathcal{J}^{\alpha}$ is given by
\begin{align}
    \mathcal{J}^{\alpha}(x,t) = \alpha c_s \phi^{\alpha} - D_{\alpha} \partial_x \phi^{\alpha} + \sum_{\beta,\gamma} G^{\alpha}_{\beta\gamma} (\phi^{\beta}\phi^{\gamma}) + \theta^{\alpha}(x,t)
\end{align}
Here, $D_{+} = D_{-} = D_s$ and the coefficient matrices $G^{\alpha}_{\beta\gamma}$ are given by Eqs.~(32) and (33) in the End Matter.
The stochastic term $\theta^{\alpha}$ represents a Gaussian white noise with correlations:
\begin{align}
    \langle \theta^{\alpha}(x,t) \theta^{\beta}(x',t')\rangle = 2 D_{\alpha} \delta_{\alpha\beta}\delta(t-t')\delta(x-x')
\end{align}

\begin{figure}[t]
\begin{center}
\includegraphics[scale=1.0]{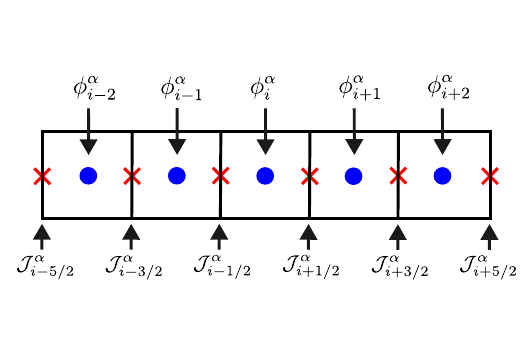}
\end{center}
\vspace{-0.5cm}
\caption{
The staggered lattice in one dimension.
The scalar fields $\phi^{\alpha}_i$ are defined at the center of each cell (indexed by integers $i=1,2,\cdots,N$), while the fluxes $\mathcal{J}^{\alpha}_{i+1/2}$ are defined at the boundaries between cells (indexed by half-integers $i+1/2$)
}
\label{supfig0}
\end{figure}
\subsection{Numerical implementation}
To numerically solve our model, we use a scheme based on the method of Garcia, Bell, and others~\cite{Garcia2024-nq}, which combines a spatial discretization on a staggered grid~\cite{BalboaUsabiaga2012-sh, Srivastava2023-nx} with a third-order stochastic Runge-Kutta (RK3) method for time integration~\cite{Delong2013-fh, Srivastava2023-nx}.

The core idea of the staggered grid is to define scalar quantities and their associated fluxes at different locations in the computational domain.
We first divide the one-dimensional space into a series of cells of width $a_{\rm uv}$.
As illustrated in Fig.~\ref{supfig0}, the scalar fields $\phi^{\alpha}_i$ are defined at the center of each cell (indexed by integers $i=1,2,\cdots,N$), while the fluxes $\mathcal{J}^{\alpha}_{i+1/2}$ are defined at the boundaries between cells (indexed by half-integers $i+1/2$).

This staggered arrangement provides a natural and robust discretization of the conservation law:
\begin{align}
    \frac{d \phi^{\alpha}_{i}}{d t} = - \frac{\mathcal{J}^{\alpha}_{i+1/2} - \mathcal{J}^{\alpha}_{i-1/2}}{a_{\rm uv}}, \label{eq:appendix_discrete_continuity}
\end{align}
where
\begin{align}
    \mathcal{J}^{\alpha}_{i+1/2} = \alpha c_s \phi^{\alpha}_{i+1/2} - D_{\alpha} (\partial_x \phi^{\alpha})_{i+1/2} + \sum_{\beta,\gamma} G^{\alpha}_{\beta\gamma} (\phi^{\beta}\phi^{\gamma})_{i+1/2} + \theta^{\alpha}_{i+1/2}. \label{eq:appendix_discrete_flux}
\end{align}
To complete the spatial discretization, the flux $\mathcal{J}^{\alpha}_{i+1/2}$ at the cell boundary must be expressed in terms of the scalar variables $\phi^{\alpha}_{i}$ at the cell centers.
We approximate the individual terms in Eq.~(\ref{eq:appendix_discrete_flux}) as follows:
\begin{enumerate}
    \item Advection term [$\alpha c_s \phi^{\alpha}_{i+1/2}$]: This term is approximated by 
    \begin{align}
        \alpha c_s \phi^{\alpha}_{i+1/2} = \alpha c_s \frac{\phi^{\alpha}_{i+1} + \phi^{\alpha}_{i}}{2}.
    \end{align}
    \item Diffusion term [$D_{\alpha} (\partial_x \phi^{\alpha})_{i+1/2}$]: This term is approximated by 
    \begin{align}
        D_{\alpha} (\partial_x \phi^{\alpha})_{i+1/2} =D_{\alpha} \frac{\phi^{\alpha}_{i+1} - \phi^{\alpha}_{i}}{a_{\rm uv}}.
    \end{align}
    \item Nonlinear term [$G^{\alpha}_{\beta\gamma} (\phi^{\beta}\phi^{\gamma})_{i+1/2}$]: This term is approximated by
    \begin{align}
        G^{\alpha}_{\beta\gamma} (\phi^{\beta}\phi^{\gamma})_{i+1/2} = G^{\alpha}_{\beta\gamma} \left(\frac{\phi^{\beta}_{i} \phi^{\gamma}_{i+1} + \phi^{\beta}_{i+1} \phi^{\gamma}_{i}}{12} + \frac{\phi^{\beta}_{i} \phi^{\gamma}_{i} + \phi^{\beta}_{i+1} \phi^{\gamma}_{i+1}}{6}\right).
    \end{align}
    Importantly, in the absence of noise, this formulation strictly conserves the quantity
    $\sum_{\alpha=+,-,0} (\phi^{\alpha})^2$~\cite{Delong2013-fh}.
\end{enumerate}

The semi-discrete equations Eq.~(\ref{eq:appendix_discrete_continuity}) form a system of stochastic ordinary differential equations.
We can write this system compactly as:
\begin{align}
\frac{d \phi^{\alpha}_{i}}{d t} = F^{\alpha}_i(\bm{\phi}) + \eta^{\alpha}_i[\bm{\theta}],
\end{align}
where $F^{\alpha}_i(\bm{\phi})$ is related to the deterministic part of the flux, $\mathcal{J}^{\alpha}_{\text{det}, i+1/2}$, (including the advection, diffusion, and nonlinear terms), and $\eta^{\alpha}_i[\bm{\theta}]$ is the stochastic part:
\begin{align}
F^{\alpha}_i[\bm{\phi}] &= - \frac{\mathcal{J}^{\alpha}_{\text{det}, i+1/2} - \mathcal{J}^{\alpha}_{\text{det}, i-1/2}}{a_{\rm uv}}, \\
\eta^{\alpha}_i[\bm{\theta}] &= - \frac{\theta^{\alpha}_{i+1/2}(t) - \theta^{\alpha}_{i-1/2}(t)}{a_{\rm uv}}.
\end{align}

We integrate this system using the RK3 scheme~\cite{Delong2013-fh, Srivastava2023-nx}.
The update from time $t_n$ to $t_{n+1}$ proceeds via two intermediate fields, $\phi_{i}^{\alpha}(t_{n+1/3})$ and $\phi_{i}^{\alpha}(t_{n+2/3})$:
\begin{align}
\phi_{i}^{\alpha}(t_{n+1/3}) &= \phi_{i}^{\alpha}(t_n) + \Delta t F^{\alpha}_i[\bm{\phi}(t_n)] + \Delta t \eta^{\alpha}_i[\bm{\theta}(t_n)], \\
\phi_{i}^{\alpha}(t_{n+2/3}) &= \frac{3}{4} \phi_{i}^{\alpha}(t_n) + \frac{1}{4} \left(\phi_{i}^{\alpha}(t_{n+1/3}) + \Delta t F^{\alpha}_i[\bm{\phi}(t_{n+1/3})] + \Delta t \eta^{\alpha}_i[\bm{\theta}(t_{n+1/3})] \right), \\
\phi_{i}^{\alpha}(t_{n+1}) &= \frac{1}{3} \phi_{i}^{\alpha}(t_n) + \frac{2}{3} \left(\phi_{i}^{\alpha}(t_{n+2/3}) + \Delta t F^{\alpha}_i[\bm{\phi}(t_{n+2/3})] + \Delta t \eta^{\alpha}_i[\bm{\theta}(t_{n+2/3})] \right).
\end{align}

The stochastic terms $\eta^{\alpha}_i[\bm{\theta}(t_n)]$, $\eta^{\alpha}_i[\bm{\theta}(t_{n+1/3})]$, and $\eta^{\alpha}_i[\bm{\theta}(t_{n+2/3})]$ are constructed as follows.
First, at the beginning of each time step, two independent arrays of Gaussian random numbers $\xi^{\alpha}_{A, i+1/2}$ and $\xi^{\alpha}_{B, i+1/2}$ are generated at each cell boundary from a standard normal distribution (mean $0$, variance $1$).
The stochastic term $\theta_{i+1/2}^{\alpha}(t_{n+k/3})$ for each stage is then constructed as
\begin{align}
\theta_{i+1/2}^{\alpha}(t_{n+k/3}) = \sqrt{\frac{2 D_{\alpha}}{a_{\rm uv} \Delta t}} \left( \xi^{\alpha}_{A, i+1/2} + \beta_k \xi^{\alpha}_{B, i+1/2} \right),
\end{align}
where the coefficients $\beta_{k}$ for the three stages are given by
\begin{align*}
  \beta_0 &= \frac{2\sqrt{2} + \sqrt{3}}{5}, \\
  \beta_1 &= \frac{-4\sqrt{2} + 3\sqrt{3}}{5}, \\
  \beta_2 &= \frac{\sqrt{2} - 2\sqrt{3}}{10}.
\end{align*}
Finally, $\eta^{\alpha}_i[\bm{\theta}(t_{n+k/3})]$ is simply the discrete divergence of the stochastic flux:
\begin{align}
    \eta^{\alpha}_i[\bm{\theta}(t_{n+k/3})] &= - \frac{\theta^{\alpha}_{i+1/2}(t_{n+k/3}) - \theta^{\alpha}_{i-1/2}(t_{n+k/3})}{a_{\rm uv}}.
\end{align}

\subsection{Parameters}
In our simulations, we use a system of units where the diffusion constant of the heat mode is set to $D_0 = 1.0$, the temperature is $k_B T = 1.0$, and the nonlinear coefficient is $\lambda_1 = 1.0$.
The numerical discretization is performed with a spatial grid spacing of $a_{\rm uv}=1.0$ and a time step of $\Delta t = 0.002$.
The remaining physical quantities—the system size $L$, the sound velocity $c_s$, the sound mode diffusion constant $D_s$, and the nonlinear coefficients $\lambda_2$, $\lambda_3$, and $\lambda_4$ —are treated as adjustable parameters in our study.

\subsection{Observation protocol}
The simulation data are acquired through a two-stage protocol.
First, we initialize each simulation with $\phi^{\alpha}(\bm{x}) = 0$, and perform a relaxation run of $4.0 \times 10^8$ steps  (corresponding to $8.0 \times 10^5$ time units) to ensure that the system reaches a non-equilibrium steady state.

After the relaxation period, a production run of $8.0 \times 10^8$ steps (corresponding to $1.6 \times 10^6$ time units) is performed.
During this run, we save the spatial profile $\phi^{\alpha}(x,t)$ every $1.0 \times 10^4$ steps.
This procedure yields a time series of $8.0 \times 10^4$ saved profiles for each simulation.
For statistical robustness, we execute 144 independent simulations with a different realization of the stochastic noise.

The primary observable is the spatio-temporal correlation function, defined as:
\begin{align}
    \Sf^\alpha_{\rm finite}(x,t) = \langle \phi^{\alpha}(x_0 + x, t_0 + t) \phi^{\alpha}(x_0,t_0)\rangle .
\end{align}
In our numerical simulations, the ensemble average $\langle \dots \rangle$ is realized by averaging over (i) all possible spatial origins $x_0$, (ii) all possible temporal origins $t_0$ available in the time-series data, and (iii) all independent simulation runs.

To perform the average over time origins, we use a block-average method.
The total time series of $8.0 \times 10^4$  saved profiles is divided into $16$ non-overlapping blocks.
Each block thus contains $N=5.0 \times 10^3$ consecutive spatial profiles.
Within a single block, we compute the correlation function for a discrete time lag $t_m = m \Delta t$ (where $\Delta t$ is the time interval between saved profiles) as
\begin{align}
    \Sf^\alpha_{\rm finite}(x,t_m) = \frac{1}{N-m} \sum_{n=1}^{N-m} \left[ \frac{1}{L} \int_0^L dx_0 \phi^\alpha(x_0+x, t_{n} + t_m) \phi^\alpha(x_0, t_n) \right].
\end{align}
Finally, we average this quantity over the $16$ blocks and the $144$ independent runs.

\section{Finite-size correction to the dynamical scaling law}
\label{sec:finite-size-correction}
In the main text, we use a modified dynamical scaling law for a finite system of size $L$:
\begin{align}
    \Sf^\alpha_{\rm finite}(x,t) = t^{-1/z} f_{\alpha}\left(\frac{x-\alpha c_s t}{t^{1/z}}\right) - C_{\alpha}.
    \label{eq:appendix_modified_scaling}
\end{align}
In this appendix, we provide its theoretical basis and present additional numerical results to support its validity.

\subsection{Theoretical verification of Eq.~(\ref{eq:appendix_modified_scaling})}
First, we theoretically derive Eq.~(\ref{eq:appendix_modified_scaling}) under certain assumptions.
In a finite system with periodic boundary conditions, the Fourier transform of the correlation function is given by
\begin{align}
    \Sf^\alpha_{\rm finite}(x,t) = \frac{1}{L} \sum_{n= -\infty}^{\infty} \Sf^\alpha_{\rm finite}(k_n,t) e^{ik_n x},
\end{align}
where the wave number $k_n$ is given by $k_n = 2 \pi n / L$ for any integer $n$.

The conservation law for the field $\phi^{\alpha}(x,t)$ imposes a global sum rule on the correlation function, which must be satisfied at all times:
\begin{align}
    \int_0^L \Sf^\alpha_{\rm finite}(x,t) dx = 0.
\end{align}
This sum rule precisely leads to the strict constraint that the zero-mode component must vanish:
\begin{align}
    \Sf^\alpha_{\rm finite}(k_0,t) = \Sf^\alpha_{\rm finite}(k = 0,t) = 0.
\end{align}
Given this constraint, the Fourier series for $\Sf^\alpha_{\rm finite}(x,t)$ can be written as a sum over all non-zero modes:
\begin{align}
    \Sf^\alpha_{\rm finite}(x,t) = \frac{1}{L} \sum_{n \neq 0} \Sf^\alpha_{\rm finite}(k_n,t) e^{ik_n x}.
    \label{eq:appendix_fourier_series_of_Sf}
\end{align}

Here, we introduce a key physical argument.
In the scaling regime where the correlation length $\xi(t)\sim t^{1/z}$ is much smaller than the system size $L$, the dynamics of short-wavelength modes ($k_n \neq 0$) are local and should not be affected by boundary effects.
Therefore, it is reasonable to approximate the behavior of these modes in the finite system by that of the corresponding modes in an infinite system:
\begin{align}
    \Sf^\alpha_{\rm finite}(k_n,t) \simeq\Sf^\alpha_{\rm infinite}(k_n,t) \quad {\rm (for \ } n \neq 0 {\rm )}.
\end{align}
Substituting this approximation into Eq.~(\ref{eq:appendix_fourier_series_of_Sf}), we obtain:
\begin{align}
        \Sf^\alpha_{\rm finite}(x,t) &= \frac{1}{L} \sum_{n \neq 0} \Sf^\alpha_{\rm infinite}(k_n,t) e^{ik_n x}, \nonumber \\
        &= \left[\frac{1}{L} \sum_{n = 0} \Sf^\alpha_{\rm infinite}(k_n,t) e^{ik_n x}\right] - \frac{1}{L} \Sf^\alpha_{\rm infinite}(k=0,t).
\end{align}
For large $L$, the term in the square bracket accurately converges to the integral that reconstructs the real-space correlation function $\Sf^\alpha_{\rm infinite}(x,t)$:
\begin{align}
    \frac{1}{L} \sum_{n = 0} \Sf^\alpha_{\rm infinite}(k_n,t) e^{ik_n x} \simeq \int_{-\infty}^{\infty} \frac{dk}{2\pi} \Sf^\alpha_{\rm infinite}(k,t) e^{ik x} = \Sf^\alpha_{\rm infinite}(x,t).
\end{align}
This leads to the crucial relationship between the finite and infinite systems:
\begin{align}
    \Sf^\alpha_{\rm finite}(x,t) = \Sf^\alpha_{\rm infinite}(x,t) - \frac{1}{L} \Sf^\alpha_{\rm infinite}(k=0,t)
\end{align}

The term $\Sf^\alpha_{\rm infinite}(k=0,t)$ is the zero-wavenumber component of the correlation function for the infinite system, which is defined by:
\begin{align}
    \Sf^\alpha_{\rm infinite}(k=0,t) = \int_{-\infty}^{\infty} \Sf^\alpha_{\rm infinite}(x,t) dx.
    \label{eq:appendix_sf_zero}
\end{align}
We now assume that for the infinite system, the standard dynamical scaling law holds:
\begin{align}
    \Sf^\alpha_{\rm infinite}(x,t) = t^{-1/z} f_{\alpha}\left(\frac{x+\alpha c_s t}{t^{1/z}}\right).
\end{align}
By substituting this scaling form into Eq.~(\ref{eq:appendix_sf_zero}), we obtain
\begin{align}
    \Sf^\alpha_{\rm infinite}(k=0,t) = \int_{-\infty}^{\infty} f_{\alpha}(u) du.
\end{align}
This implies that $\Sf^\alpha_{\rm infinite}(k=0,t)$ is a time-independent constant that depends only on the shape of the universal function $f_{\alpha}$.

Thus, we have shown that the modified dynamical scaling law Eq.~(\ref{eq:appendix_modified_scaling}) holds for finite systems, and the constant offset is given by
\begin{align}
    C_{\alpha} = \frac{1}{L}\int_{-\infty}^{\infty} f_{\alpha}(u) du.
    \label{eq:appendix_relation_C_L}
\end{align}

\begin{figure}[t]
\begin{center}
\includegraphics[scale=1.0]{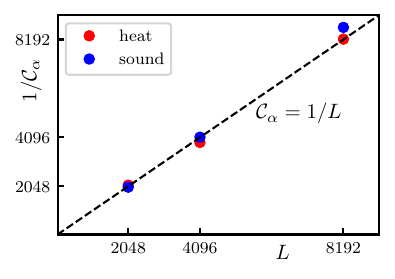}
\end{center}
\vspace{-0.5cm}
\caption{
The inverse of the finite-size correction term, $1/C_{\alpha}$, as a function of the system size $L$, showing excellent agreement with the relation $1/C_{\alpha} = L$ (dashed line).
}
\label{supfig1}
\end{figure}
\subsection{Numerical verification of Eq.~(\ref{eq:appendix_modified_scaling})}

To validate the modified scaling law Eq.~(\ref{eq:appendix_modified_scaling}), we numerically verify the theoretical prediction, $C_{\alpha} \propto 1/L$, given in Eq.~(\ref{eq:appendix_relation_C_L}).

To this end, we focus on the correlation function at $x=0$ and $t=0$, $\Sf^\alpha_{\rm finite}(x=0, t=0)$.
For an infinite system, this quantity is exactly calculated as
\begin{align}
\Sf^\alpha_{\rm infinite}(x=0, t=0) = 1.
\end{align}
We then  obtain
\begin{align}
\Sf^\alpha_{\rm finite}(x=0, t=0) \simeq 1 - C_{\alpha}.
\end{align}
This allows $C_{\alpha}$ to be computed directly from the finite-system data:
\begin{align}
C_{\alpha} = 1 - \Sf^\alpha_{\rm finite}(x=0, t=0).
\label{eq:appendix_C_alpha_calc}
\end{align}

In Fig.~\ref{supfig1}, we present the inverse of the correction term, $1/C_{\alpha}$, plotted as a function of the system size $L$.
The data points clearly follow a linear relationship, $1/C_{\alpha} \propto L$.
This numerical result is consistent with the theoretical prediction $C_{\alpha} \propto 1/L$ derived in Eq.~(\ref{eq:appendix_relation_C_L}), supporting the validity of the proposed finite-size correction.

\section{Parameter-independence of numerical results}
In the main text, we presented the numerical results for the dynamical scaling of $\Sf^{\alpha}(x,t)$ using the following parameter set:
\begin{align}
    & L = 8192, \\
    & \cs = 0.1, \nonumber \\
    & D_s = D_h = 1.0, \nonumber \\
    & \lambda_1 = \lambda_2 = \lambda_3 = \lambda_4 = 1.0. \nonumber
\end{align}
Furthermore, in the End Matter, we demonstrated that the choice of the sound velocity $\cs$ does not alter the dynamical scaling behavior.
In this section, we provide additional data to demonstrate that our conclusions are robust against changes in the other parameters, specifically the system size $L$, the diffusion constant $D_s$, and the nonlinear coupling constants $\lambda_i$ ($i=1, 2, 3, 4$).

\begin{figure*}[t]
\begin{center}
\includegraphics[scale=0.9]{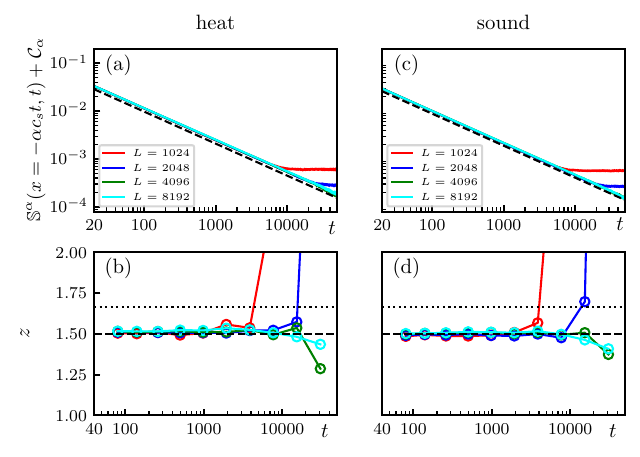}
\end{center}
\vspace{-0.5cm}
\caption{
The change in $\Sf^\alpha(x=-\alpha c_s t,t)$ for different system sizes.
Top panels: Log-log plots of time correlation $\Sf^\alpha(x=-\alpha c_s t,t)$ for (a) heat and (c) sound modes.
Bottom panels: Extracted dynamical exponent $z$ for (b) heat and (d) sound modes.
$L=1024,2048,4096,8192$. $c_s=0.1$. Other parameters = 1.0.
Dashed lines indicate the KPZ prediction ($\propto t^{-2/3}$ and $z=3/2$).
}
\label{supfig3}
\end{figure*}
\subsection{System size}
We first examine the effect of the system size $L$ on the time correlation $\Sf^\alpha(x=0,t)$ in Fig.~\ref{supfig3}.
As discussed in Sec.~\ref{sec:finite-size-correction}, the finite-size effect arising from the conservation law can be corrected by subtracting $C_{\alpha}$.
However, a residual finite-size effect persists due to the periodic boundary conditions.
This boundary effect becomes significant at a characteristic time $t_c \approx L/(2c_s)$, which corresponds to the time required for counter-propagating sound waves to meet.
Consequently, as shown in Figs.~\ref{supfig3}(a) and (c), the correlation deviates from the power-law decay and saturates at long times.
The bottom panels [Figs.~\ref{supfig3}(b) and (d)] show the extracted exponent $z$.
As $L$ increases, the time window in which $z=3/2$ (i.e., $\propto t^{-2/3}$) widens.
This confirms that the RG prediction $z=3/2$ is recovered in the $L \to \infty$ limit.
The size $L=8192$ used in the main text provides a characteristic time of $t_c \approx 40960$ (for $c_s=0.1$), which is sufficiently large to observe the scaling behavior before the finite-size effect becomes dominant.

\begin{figure*}[t]
\begin{center}
\includegraphics[scale=0.9]{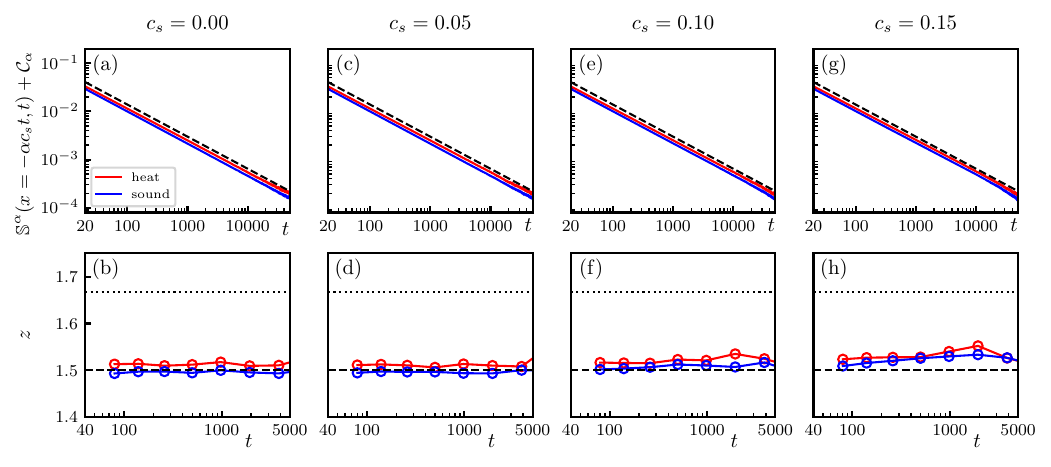}
\end{center}
\vspace{-0.5cm}
\caption{
Robustness of KPZ scaling ($z=3/2$) against varying sound velocity $c_s$.
Top panels (a,c,e,g): Log-log plots of time correlation $\Sf^\alpha(x=-\alpha c_s t,t)$.
Bottom panels (b,d,f,h): Extracted dynamical exponent $z$.
Tested $c_s = 0.0, 0.05, 0.1, 0.15$. Other parameters = $1.0$.
Dashed lines indicate the KPZ prediction ($\propto t^{-2/3}$ and $z=3/2$).
}
\label{supfig4}
\end{figure*}
\begin{figure*}[t]
\begin{center}
\includegraphics[scale=0.9]{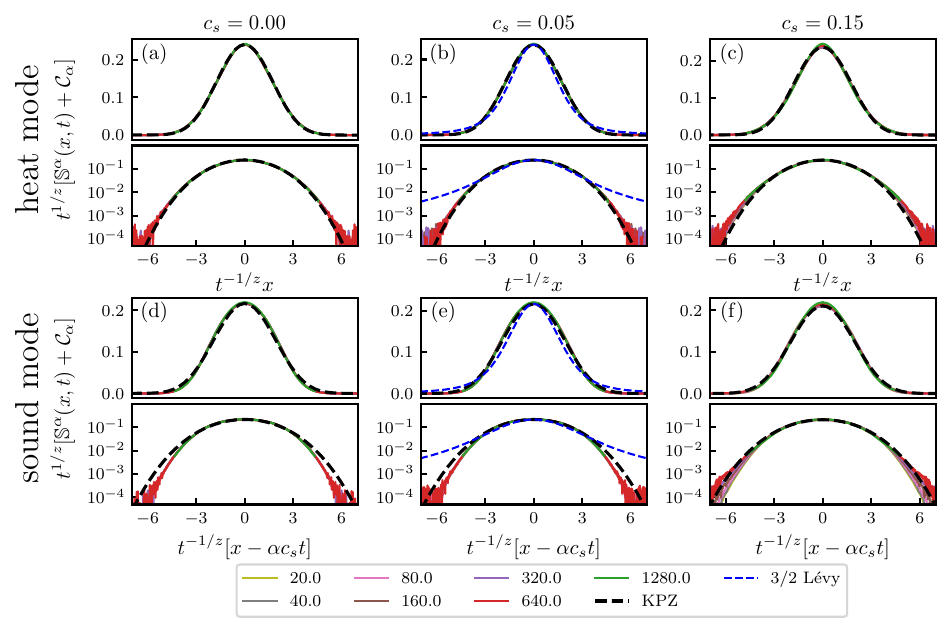}
\end{center}
\vspace{-0.5cm}
\caption{
Robustness of the dynamical scaling against varying sound velocity $c_s$, shown via data collapse.
Results for heat (a-c) and sound (d-f) modes.
Parameter sets: (a,d) $c_s = 0.00$; (b,e) $c_s = 0.05$; (c,f) $c_s = 0.15$.
Other parameters = $1.0$
Black dashed curve: theoretical KPZ function, blue dashed curves: $3/2$-Lévy distribution.
}
\label{supfig5}
\end{figure*}
\subsection{Sound velocity}
We provide data in Fig.~\ref{supfig4} to support the robustness of the dynamical scaling against the sound velocity $\cs$.
This figure focuses on the time correlation function $\Sf^\alpha(x=0,t)$.
The top panels [Fig.~\ref{supfig4}(a, c, e, g)] display log-log plots of $\Sf^\alpha(x=0,t)$ versus time $t$ for four different values of $\cs$.
In all cases, the data exhibit a power-law decay consistent with the RG prediction $z=3/2$ (i.e., $\propto t^{-2/3}$), as indicated by the dashed lines.
More quantitatively, the bottom panels [Fig.~\ref{supfig4}(b, d, f, h)] show the dynamical exponent $z$ extracted from fitting these decays.
These results, which correspond to the analysis shown in Fig.~1(b) of the main text, confirm that the extracted exponent $z$ is in excellent agreement with the theoretical KPZ value $z=3/2$ for all tested sound velocities.

In addition, we test the robustness of the scaling functions against $c_s$ in Fig.~\ref{supfig5}.
This figure shows the data collapse for both heat (a-c) and sound (d-f) modes, using three different $c_s$ values.
In all cases, the numerical data (colored dots) collapse onto a single curve and show excellent agreement with the theoretical scaling functions for the KPZ equation (black dashed curve).
This result confirms that the universal scaling behavior is independent of the specific value of the sound velocity.

\begin{figure*}[t]
\begin{center}
\includegraphics[scale=0.9]{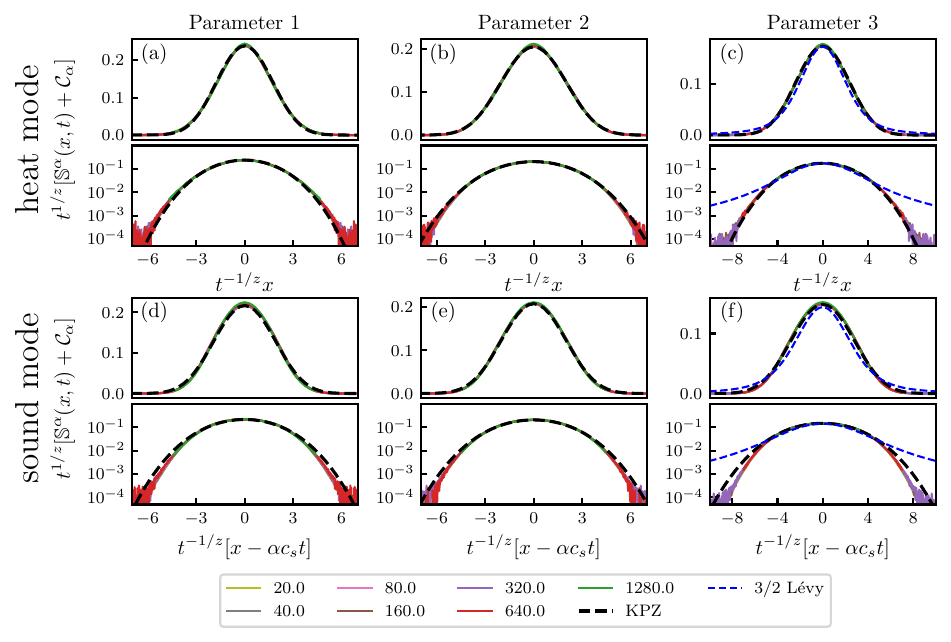}
\end{center}
\vspace{-0.5cm}
\caption{
Robustness of the dynamical scaling against varying nonlinear parameters $\lambda_i$, shown via data collapse.
Results for heat (a-c) and sound (d-f) modes.
Parameter sets: (a,d) $\{\lambda_1, \lambda_2, \lambda_3, \lambda_4\} = \{1.0, 1.0, 1.0, 1.0\}$; (b,e) $\{1.0, 1.6, 1.1, 0.6\}$; (c,f) $\{1.0, 2.5, 1.7, 1.3\}$.
Fixed $c_s=0.10$. Other parameters = $1.0$.
Black dashed curve: theoretical KPZ function, blue dashed curves: $3/2$-Lévy distribution.
}
\label{supfig6}
\end{figure*}
\subsection{Nonlinear parameters}
Next, we examine the dependence on the nonlinear coupling constants $\lambda_i$. Figure~\ref{supfig6} illustrates the dynamical scaling of $\Sf^{\alpha}(x,t)$ for three different sets of $\lambda_i$ values, as specified in the figure caption.
The results are presented for both the heat mode [Fig.~\ref{supfig6}(a-c)] and the sound mode [Fig.~\ref{supfig6}(d-f)].
All other parameters are fixed, with $\cs=0.10$. 
The panels show the data collapse of $\Sf^{\alpha}(x,t)$ plotted as a function of the scaled variables $t^{-1/z}x$ and $t^{1/z}[\Sf^{\alpha}(x + \alpha \cs t,t) + C_{\alpha}]$ with the best-fit values $z$.
In all tested cases, the numerical data collapse well onto a single curve and show good agreement with the theoretical KPZ scaling function (black dashed curve).
This demonstrates that the KPZ scaling is robust against variations in the nonlinear coupling constants.

\begin{figure*}[t]
\begin{center}
\includegraphics[scale=0.9]{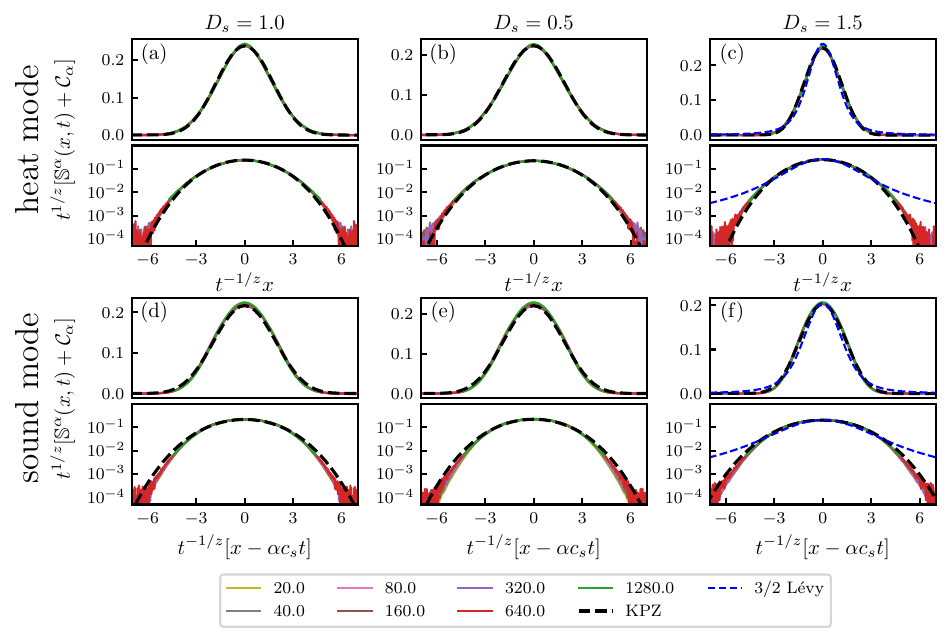}
\end{center}
\vspace{-0.5cm}
\caption{
Robustness of the dynamical scaling against varying diffusion constant $D_s$, shown via data collapse.
Results for heat (a-c) and sound (d-f) modes.
Parameters: (a, d) $D_s=1.00$; (b, e) $0.50$; (c, f) $1.50$.
Fixed $c_s=0.10$. Other parameters = $1.0$.
Black dashed curve: theoretical KPZ function, blue dashed curves: $3/2$-Lévy distribution.
}
\label{supfig7}
\end{figure*}
\subsection{Diffusion constant}
Finally, we test the robustness of the scaling results against variations in the diffusion constant of the sound mode, $D_s$.
Figure~\ref{supfig7} presents the results for three different values of $D_s$ ($1.00$, $0.50$, and $1.50$), while $\cs$ is fixed at $0.10$ and all other parameters are set to $1.0$.
The top panels display the spatial profiles, while the bottom panels show the corresponding data collapse.
Similar to the results for the nonlinear parameters, in all tested cases, the numerical data collapse well onto a single curve and show good agreement with the theoretical KPZ scaling function (black dashed curve).
This confirms that the observed KPZ dynamical scaling is independent of the specific choice of $D_s$.

\section{Remarks on the original nonlinear fluctuating hydrodynamics}
We reexamine the nonlinear fluctuating hydrodynamics (NFH) theory proposed in Ref.~\cite{spohn2014nonlinear}, which can be written in the same form as Eq.~(\ref{EOM original}) but with the following coefficients:
\begin{align}
A &= \begin{pmatrix}
     0 & -1 & 0\\
    \partial_l P & 0 & \partial_e P\\
    0 & P & 0
\end{pmatrix}, \label{Spohn A matrix} \\
D 
&= \begin{pmatrix}
0 & 0 & 0 \\
0 & D_u & 0 \\
\tilde{D}_e & 0 & D_e
\end{pmatrix}, \\
H^2 &=\frac{1}{2} \begin{pmatrix}
\partial_l^2 P & 0 & \partial_l\partial_e P\\
0 & \partial_e P & 0 \\
\partial_l\partial_e P& 0 & \partial_e^2 P
\end{pmatrix},\\
H^3 &= \frac{1}{2}\begin{pmatrix}
0 & \partial_l\partial_e P & 0 \\
\partial_l\partial_e P & 0 & \partial_lP \\
0 & \partial_lP & 0
\end{pmatrix}, 
\end{align}
and $H^1=\mathcal{O}$ where $\mathcal{O}$ is the zero matrix.
Here, $P$ is the pressure and  $D_u$, $D_e$, and $\tilde{D}_e$  are the diffusion constants.
The coefficients $B$  are determined to satisfy the equilibrium condition ($\ref{eq:FDT1}$).
We note that all components of $H$ are expressed in terms of thermodynamic derivatives.
However, as we show below, this formulation fails to satisfy the equilibrium condition~(\ref{potential condition}) for general equilibrium states. 

By imposing Eq.~(\ref{potential condition}) on $H$, 
we obtain the following constraints on the thermodynamic derivatives:
\begin{align}
    \partial_l^2 P &= a_0 \partial_l P,\quad \partial_e^2P = a_3 \partial_eP, \\
    \partial_l\partial_e P &= a_3 \partial_lP, \quad 
    \partial_l\partial_e P = a_0 \partial_eP. \label{eq:potential cond Spohn}
\end{align}
These additional constraints on the thermodynamic quantities indicate that the original NFH theory proposed in Ref.~\cite{spohn2014nonlinear} does not satisfy the thermalization condition for general equilibrium states.
These relations cannot be satisfied simultaneously for general $P(l,e)$, indicating that the formulation is physically inconsistent.

Furthermore, these constraints lead to an unphysical situation, as shown below.  
From Eq.~(\ref{eq:potential cond Spohn}), we obtain
\begin{align}
    a_3\,\partial_l P = a_0\,\partial_e P. \label{eq:pot cond Spohn 2}
\end{align}
To  express the thermodynamic derivatives in terms of the parameters $a_i$,  
we next follow the formulation adopted in Ref.~\cite{spohn2014nonlinear}.
Specifically, we summarize its construction based on the local Gibbs distribution for the conserved quantities written in terms of the microscopic variables $r_i = q_{i+1} - q_i$, $p_i$, and $\epsilon_i = p_i^2/(2m) + V(r_i)$.  
We collectively denote these by $\mathcal{U}_{a,i} = (r_i, p_i, \epsilon_i)$, and write the distribution as
\begin{align}
    \Pfp_{\mathrm{LG}}[\mathcal{U}_{a,i}] 
    = \frac{1}{Z_{\mathrm{LG}}} 
    \exp\biggl[-\sum_{a,i} \lambda^a_{i}\, \mathcal{U}_{a,i}\biggr],
\end{align}
where $\lambda^a_i$ are the local intensive variables.  
Note that this definition of the hydrodynamic fields---particularly the stretch---differs from ours, but we follow this convention here for consistency with Ref.~\cite{spohn2014nonlinear}.
At global equilibrium, they take constant values
\begin{align}
    \bm{\lambda}_{\mathrm{eq}} = (\beta P,\, 0,\, \beta),
\end{align}
with $\beta$ being the inverse temperature. 
The hydrodynamic fields $u_a(x)$ in Ref.~\cite{spohn2014nonlinear} are then defined  as the local deviations of these conserved quantities from their equilibrium averages,
\begin{align}
    u_a(x)
    := \langle \mathcal{U}_{a,i}\rangle_{\mathrm{LG}}
    - \langle \mathcal{U}_{a,i}\rangle_{\mathrm{eq}}.
\end{align}
The pressure $P$ is regarded as a thermodynamic function of the conserved quantities $u_a$, 
and its derivatives are related to the linear response of the intensive variables. 
Specifically, using
\begin{align}
    \frac{\partial \lambda^a}{\partial u_b} = -\Co^{-1}_{ab}, \label{eq:dlam_du}
\end{align}
which follows directly from the definition of the covariance matrix in the local Gibbs distribution, we find
\begin{align}
    \partial_l P 
    &= \frac{1}{\beta}\biggl(\frac{\partial(\beta P)}{\partial l} - P\, \frac{\partial \beta}{\partial l}\biggr), \nonumber \\
    &= \frac{1}{\beta}\biggl(\frac{\partial \lambda^1}{\partial u_1} - P\, \frac{\partial \lambda^3}{\partial u_1} \biggr), \nonumber \\
    &= \beta^{-1}\!\left(-\Co^{-1}_{11} + P\, \Co^{-1}_{13}\right), \nonumber \\
    &= \beta^{-1}\!\left(a_0 P - a_1\right), \\
    \partial_e P 
    &= \beta^{-1}\!\left(a_3 P - a_0\right).
\end{align}
Using these expressions, the constraint~(\ref{eq:pot cond Spohn 2}) reduces to
\begin{align}
    a_0^2 - a_1 a_3 = 0.
\end{align}
This relation implies a singular equilibrium distribution with $\det \Co^{-1} = 0$.
Therefore, the NFH theory proposed in Ref.~\cite{spohn2014nonlinear} satisfies the equilibrium condition only in a singular, hence unphysical, state.
We note that the $A$ matrix in Ref.~\cite{spohn2014nonlinear} satisfies the equilibrium condition~(\ref{eq:FDT2}) without any additional constraints and coincides with that obtained in our derivation.

Furthermore, the $G$ matrix in Ref.~\cite{spohn2014nonlinear} is characterized by the following symmetry relations:
\begin{align}
G^\alpha_{\beta \gamma}&=G^\alpha_{\gamma \beta}, \quad
G^\sigma_{\beta \gamma}=-G^{-\sigma}_{-\beta -\gamma}, \quad
G^+_{0 +}=G^+_{0 -}, \nonumber\\
G^0_{++}&=-G^0_{--}, \quad 
G^0_{\beta\gamma}=0 \; \text{(otherwise)},
\end{align}
where $\sigma=\pm$. The parametrized expressions are given by
\begin{align}
    G^+&=\begin{pmatrix}
        \nu_1 & \nu_2 & \nu_3\\
        \nu_2 & \nu_4 & \nu_2  \\
        \nu_3 & \nu_2  &\nu_5
    \end{pmatrix}, \,~
       G^-=\begin{pmatrix}
        -\nu_5 &  -\nu_2 & -\nu_3\\
        -\nu_2  & -\nu_4 & -\nu_2 \\
        -\nu_3 &-\nu_2& -\nu_1
    \end{pmatrix}, \, ~
        G^0=\begin{pmatrix}
        \nu_0 & 0 & 0\\
         0 & 0 & 0 \\
        0 & 0& -\nu_0
    \end{pmatrix}. 
\end{align}
Using this $G$ in the RG analysis, however, we did not obtain a nontrivial fixed point other than the trivial EW one when solving the RG equations numerically with Mathematica.
This result indicates that the NFH proposed in Ref.~\cite{spohn2014nonlinear} fails to capture the nontrivial fixed-point structure responsible for the KPZ universality class. 
We also performed extensive numerical computations for this parameter setting and found that most parameter choices led to numerically unstable time evolutions. 
Even for numerically stable parameter regimes, compared to our model, it remains unclear whether both the heat and sound modes exhibit the scaling behavior~\cite{DR}.

\end{document}